\documentclass[aps,twocolumn,superscriptaddress]{revtex4-2}
\usepackage{newtxtext}
\usepackage{lipsum}

\usepackage{amsmath}
\usepackage{amssymb}
\usepackage{mathtools}
\usepackage{empheq}
\usepackage{graphicx}
\usepackage{xcolor}
\usepackage{bm}
\usepackage{mathrsfs}
\usepackage{colortbl}
\usepackage[version=4]{mhchem}

\definecolor{myred}{HTML}{E1558A}
\definecolor{myblue}{HTML}{318294}
\definecolor{myskyblue}{HTML}{3FC3E0}

\newcommand{\add}[1]{#1}

\usepackage[hidelinks]{hyperref}
\hypersetup{
  colorlinks   = true, 
  urlcolor     = blue, 
  linkcolor    = myred, 
  citecolor   = myblue 
}

\newcommand{\cip}[2]{\left\langle #1, #2 \right\rangle}

\DeclareMathOperator{\csch}{csch}

\begin{document}
\title{Infinite variety of thermodynamic speed limits with general activities}
\author{Ryuna Nagayama}
\email{ryuna.nagayama@ubi.s.u-tokyo.ac.jp}
\affiliation{Department of Physics, The University of Tokyo, 7-3-1 Hongo, Bunkyo-ku, Tokyo 113-0033, Japan}
\author{Kohei Yoshimura}
\affiliation{Department of Physics, The University of Tokyo, 7-3-1 Hongo, Bunkyo-ku, Tokyo 113-0033, Japan}
\author{Sosuke Ito}
\affiliation{Department of Physics, The University of Tokyo, 7-3-1 Hongo, Bunkyo-ku, Tokyo 113-0033, Japan}
\affiliation{Universal Biology Institute, The University of Tokyo, 7-3-1 Hongo, Bunkyo-ku, Tokyo 113-0033, Japan}

\begin{abstract}
    Activity, which represents the kinetic property of dynamics, plays a central role in obtaining thermodynamic speed limits (TSLs). In this paper, we discuss a unified framework that provides the existing TSLs based on different activities such as dynamical activity and dynamical state mobility. \add{
    This unification is based on generalized means that include standard means such as the arithmetic, logarithmic, and geometric means, the first two of which respectively correspond to the dynamical activity and the dynamical state mobility.
    } We also derive an infinite variety of TSLs for Markov jump processes and deterministic chemical reaction networks using different activities. The lower bound on the entropy production given by each TSL provides the minimum dissipation achievable by a conservative force. We numerically and analytically discuss the tightness of the lower bounds on the EPR in the various TSLs.
\end{abstract}


\maketitle

\section{Introduction}

One of the fundamental aims of nonequilibrium thermodynamics is to discover the laws governing dissipation, specifically the entropy production (EP) or its rate, the entropy production rate (EPR). The oldest and best known example is the second law of thermodynamics, which states that the EP is nonnegative. 
Recently, with the development of stochastic thermodynamics~\cite{sekimoto2010stochastic, seifert2012stochastic}, more refined laws applicable to Markov processes have been discovered. A prominent example is the thermodynamic speed limits (TSLs), which relate the speed of time evolution to the EP~\cite{aurell2012refined, shiraishi2018speed, chen2019stochastic, nakazato2021geometrical}. The TSLs have been shown to hold universally for various systems, including deterministic chemical reaction networks (CRNs)~\cite{yoshimura2021thermodynamic,yoshimura2023housekeeping,van2023topological,kolchinsky2022information},  deterministic reaction-diffusion systems~\cite{nagayama2023geometric}, much like the second law of thermodynamics.

For Markov jump processes (MJPs), the dynamical activity, which determines the kinetic intensity of the system, plays a crucial role in the TSLs. A typical TSL for MJPs provides a lower bound on the EP by rescaling the square of the transition speed by the dynamical activity~\cite{Maes2008,shiraishi2018speed}. To date, various efforts have been made to refine the relations between these three elements: the transition speed, the dynamical activity, and the EP. The variety of TSLs is based on the different approaches such as refining the functional forms that appear in TSLs~\cite{lee2022speed,Vo2022,Falasco2022,kolchinsky2022information,delvenne2024thermokinetic}, measuring the speed with the 1-Wasserstein distance and the 2-Wasserstein distance from optimal transport theory~\cite{villani2009optimal,maas2011gradient,santambrogio2015optimal, peyre2019computational} instead of the simple total variation distance~\cite{dechant2022minimum,yoshimura2023housekeeping,van2023thermodynamic,delvenne2024thermokinetic,nagayama2023geometric}, and using the Hatano-Sasa excess EP~\cite{hatano2001steady} or other generalizations of the excess EP, which is the portion of the EP that inherently affects the time evolution, instead of the total EP~\cite{shiraishi2018speed,vo2020unified,kolchinsky2022information, yoshimura2023housekeeping}.

For the TSLs based on the Wasserstein distances, alternative quantities are used instead of the dynamical activity to measure the kinetic intensity of the system: the dynamical state mobility~\cite{van2023thermodynamic} and the edgewise Onsager coefficient~\cite{yoshimura2023housekeeping}. The difference between the dynamical activity and these alternative quantities can be understood in terms of means: the dynamical activity is twice the sum of the arithmetic mean of the forward and reverse jump rates for all transitions, whereas the dynamical state mobility is the sum of the logarithmic mean of these rates. The edgewise Onsager coefficient is also introduced as the logarithmic mean of these rates for each transition. This mean perspective suggests that it is possible to derive distinct TSLs by altering the mean used to measure the kinetic intensity.

Indeed, in nonequilibrium thermodynamics for MJPs and CRNs, some activities have been developed based on various means different from the arithmetic and logarithmic means. For example, the geometric mean of bidirectional fluxes has been found in macroscopic fluctuation theory~\cite{mielke2014relation,mielke2017non,kaiser2018canonical,Maes2008,Barato2015}. This activity based on the geometric mean provides a lower bound of the EPR~\cite{maes2017frenetic} and helps to decompose the EPR~\cite{kobayashi2022hessian,kobayashi2023information}. Furthermore, activities based on more general means have been proposed to handle relaxation toward equilibrium mathematically~\cite{peletier2022jump}. However, it remains unclear whether these general activities can similarly lead to the derivation of the TSLs as those based on the arithmetic and logarithmic means.

In this paper, we derive an infinite variety of TSLs for MJPs and deterministic CRNs using general activities. We also use the 1-Wasserstein distance and its extension to CRNs~\cite{nagayama2023geometric,kolchinsky2022information} to measure the speed of the time evolution. Our TSLs encompass previously established forms~\cite{dechant2022minimum,van2023thermodynamic}. We prove that a broad class of means, such as the Stolarsky mean containing infinitely many types of means~\cite{stolarsky1975generalizations,cisbani1938contributi,tobey1967two}, can be used as an activity. Because we can derive different TSLs for each choice of means, we obtain an infinite variety of TSLs. In nonstationary states, we numerically confirm that the tightness of the lower bounds on the EPR can vary in these different TSLs and that there is no apparent hierarchy for the tightness of the lower bounds in general. We show analytically that a hierarchy exists for the two specific means, i.e., maximum and minimum, and that a hierarchy also exists in the low-speed regime. We also numerically compare the proposed TSLs with the existing TSLs based on the 2-Wasserstein distance as lower bounds on the excess EPR~\cite{yoshimura2023housekeeping,nagayama2023geometric,kolchinsky2022information}. In contrast to the EPR, we numerically confirm that the TSL for the excess EPR does not hold for some choice of mean.


We also reveal that each TSL yields minimum dissipation under different conditions depending on the mean employed as the activity. This unifies the two previously proposed approaches, one based on the arithmetic mean~\cite{dechant2022minimum} and the other on the logarithmic mean~\cite{van2023thermodynamic}. We show that a conservative force and a time-independent current can achieve this minimum dissipation, regardless of the mean used as the activity. For MJPs, whether a conservative force achieves minimum dissipation depends on the conditions for the minimization~\cite{remlein2021optimality,ilker2022shortcuts,dechant2022minimum,van2023thermodynamic,yoshimura2023housekeeping}. Our results provide a general class of conditions under which a conservative force can achieve minimum dissipation.


\section{Dynamics and thermodynamics on networks and chemical reaction networks}

In this study, we consider Markov jump processes (MJPs) and deterministic chemical reaction networks (CRNs). Based on the mathematical analogy~\cite{ge2016mesoscopic,rao2016nonequilibrium,yoshimura2023housekeeping}, we can treat these different systems in the same way as shown below.

\subsection{Dynamics}

\subsubsection{Markov jump process}

Firstly, we consider a stochastic system consisting of $N_{S}$ microstates without odd variables, which is described by an MJP. We index the microstates by $\alpha\in\mathcal{S}$, where $\mathcal{S}$ denotes the index set of the microstates $\{1,2,\cdots,N_{S}\}$. We assume that the system is coupled with $N_{R}$ heat reservoirs. We also define the index set of the thermodynamic reservoirs as $\mathcal{R}\coloneqq\{1,2,\cdots,N_{R}\}$ and index the reservoirs by $\nu\in\mathcal{R}$.

We let the column vector $x(t)=(x_1(t),x_2(t),\cdots,x_{N_{S}}(t))^{\top}$ denote the probability distribution on the microstates at time $t$. We assume that the probability distribution satisfies $x_{\alpha}(t)>0$ for all $\alpha\in\mathcal{S}$ and $\sum_{\alpha\in\mathcal{S}}x_{\alpha}(t)=1$.

The probability distribution $x(t)$ evolves according to the following linear master equation,
\begin{align}
    d_tx_{\alpha}(t)=\sum_{\nu\in\mathcal{R}}\sum_{\beta\in\mathcal{S}}R_{\alpha\beta}^{(\nu)}(t)x_{\beta}(t),
    \label{linear master eq}
\end{align}
where $d_t$ denotes the time derivative $d/dt$, and $R_{\alpha\beta}^{(\nu)}(t)$ is the transition rate from microstate $\beta$ to microstate $\alpha$ induced by reservoir $\nu$ at time $t$. We will omit the argument $t$ if we do not focus on the time dependence. We make the following four assumptions on the transition rates: (i) $R_{\alpha\beta}^{(\nu)}$ is nonnegative if $\alpha\neq\beta$, (ii) $R_{\alpha\beta}^{(\nu)}$ is positive if and only if $R_{\beta\alpha}^{(\nu)}$ is positive, (iii) $R_{\alpha\beta}^{(\nu)}$ is always positive if it is positive at the initial time, and (iv) $R_{\alpha\alpha}^{(\nu)}$ satisfies $R_{\alpha\alpha}^{(\nu)}=-\sum_{\beta\in\mathcal{S}\setminus\{\alpha\}}R_{\beta\alpha}^{(\nu)}\leq0$ for all $\alpha\in\mathcal{S}$ and $\nu\in\mathcal{R}$. Assumption (iv) ensures the conservation of probability, that is, $d_t\sum_{\alpha\in\mathcal{S}}x_{\alpha}=0$.

To simplify the notation, we introduce the \textit{directed edges} corresponding to the transitions as 
\begin{align}
    (\beta\to\alpha;\nu),
\end{align}
for $\alpha,\beta\in\mathcal{S}$ and $\nu\in\mathcal{R}$ that satisfy $\alpha>\beta$ and $R_{\alpha\beta}^{(\nu)}>0$. Due to assumption (iii) for the transition rates, we can define the directed edges independently of time. We refer to $\beta$, $\alpha$, and $\nu$ as the start point, the target point, and the corresponding reservoir of the directed edge $(\beta\to\alpha;\nu)$. We define the index set of the directed edges as $\mathcal{E}\coloneqq\{1,2,\cdots,N_{E}\}$ with the number of the directed edges $N_{E}$. We index the directed edges by $e\in\mathcal{E}$. We also let $\mathrm{s}(e)$, $\mathrm{t}(e)$, and $\mathrm{r}(e)$ denote the start point, the target point, and the corresponding reservoir of edge $e$, respectively. This enables us to represent edge $e$ as $(\mathrm{s}(e)\to\mathrm{t}(e);\mathrm{r}(e))$.

Using the directed edges, we define the forward and reverse fluxes on edge $e\in\mathcal{E}$ as
\begin{align}
    J^{+}_e(x,t)&\coloneqq R_{\mathrm{t}(e)\mathrm{s}(e)}^{(\mathrm{r}(e))}(t)x_{\mathrm{s}(e)}(t),\notag\\J^{-}_e(x,t)&\coloneqq R_{\mathrm{s}(e)\mathrm{t}(e)}^{(\mathrm{r}(e))}(t)x_{\mathrm{t}(e)}(t).
\end{align}
Here, $J_e^{+}(x;t)dt$ gives the expected frequency of jump $e$ in a short time interval $dt$, while $J_e^{-}(x;t)dt$ amounts to that of the inverse jump, which can be expressed as $(\mathrm{t}(e)\to \mathrm{s}(e);\mathrm{r}(e))$. Note that all forward and reverse fluxes are positive due to the assumptions on the probability distributions and the transition rates. The forward and reverse fluxes let us define the (net) current along edge $e$ as
\begin{align}
    J_e(x,t)\coloneqq J^{+}_e(x,t)-J^{-}_e(x,t).
\end{align}
We also introduce the column vectors of the fluxes and the currents as $J^{\pm}(x,t)\coloneqq(J^{\pm}_1(x,t),J^{\pm}_2(x,t),\cdots,J^{\pm}_{N_E}(x,t))^{\top}$ and $J(x,t)\coloneqq(J_1(x,t),J_2(x,t),\cdots,J_{N_E}(x,t))^{\top}$. In the following, we will omit the arguments $x$ and $t$ to write $J(x)$, $J(t)$, or $J$, if we do not focus on the corresponding dependencies.

We can consider a directed graph with $\mathcal{S}$ as the set of vertices and $\mathcal{E}$ as the set of directed edges. This graph is characterized by an $N_{E}\times N_{S}$ matrix $\nabla$, whose element is defined as
\begin{align}
    \nabla_{e\alpha}\coloneqq\delta_{\alpha\mathrm{t}(e)}-\delta_{\alpha\mathrm{s}(e)}.
\end{align}
for $e\in\mathcal{E}$ and $\alpha\in\mathcal{S}$. We refer to this matrix as the \textit{gradient matrix} because it acts as the gradient operator. The transpose of the gradient matrix $\nabla^{\top}$ can be interpreted as the negative divergence operator. For example, we can rewrite the linear master equation~\eqref{linear master eq} as the following form:
\begin{align}
    d_tx=\nabla^{\top}J=\nabla^{\top}(J^{+}-J^{-}).
    \label{continuity eq}
\end{align}
If we regard $\nabla^{\top}$ as the negative divergence, we can interpret this equation as a discrete continuity equation.
We also note that the matrix $\nabla^{\top}$ is often called the \textit{incidence matrix} of the directed graph~\cite{bondy2008graph}.

\subsubsection{Chemical reaction network}
Here, we use the same symbols as those used in the case of the MJP to indicate quantities in the CRN that play common roles. 

We consider a CRN consisting of $N_{S}$ internal species and $N_{E}$ reversible reactions~\cite{feinberg2019foundations} at homogeneous temperature. We ignore external species that are exchanged with the outside of the system since they do not affect our results. We let $\mathcal{S}$ and $\mathcal{E}$ denote the index sets of the internal species $\{1,\cdots,N_{S}\}$ and the reversible reactions $\{1,\cdots,N_{E}\}$, respectively. We also let $X_{\alpha}$ denote species $\alpha\in\mathcal{S}$.    

The CRN is characterized by the number of molecules of species $\alpha$ consumed (produced) through reaction $e$, denoted by $n_{\alpha e}^+$ ($n_{\alpha e}^-$). Letting $X_{\alpha}$ denote species $\alpha\in\mathcal{S}$, reaction $e$ is given by
\begin{align}
    \ce{$\sum_{\alpha\in\mathcal{S}}n_{\alpha e}^+X_{\alpha}$ <=> $\sum_{\alpha\in\mathcal{S}}n_{\alpha e}^-X_{\alpha}$}.
\end{align}
Using $n_{\alpha e}^{\pm}$, we define the $N_{E}\times N_{S}$ \textit{gradient matrix} of this CRN $\nabla$ as
\begin{align}
    \nabla_{e\alpha}\coloneqq n_{\alpha e}^{-}-n_{\alpha e}^+,
    \label{stoichiometric matrix}
\end{align}
for $e\in\mathcal{E}$ and $\alpha\in\mathcal{S}$. The element $\nabla_{e\alpha}$ represents the net increase of the molecules of $X_{\alpha}$ through reaction $e$. The transpose of the gradient matrix $\nabla^{\top}$ is known as the stoichiometric matrix. As in the case of the MJP, the gradient matrix and its transpose act as the gradient operator and the negative divergence operator, respectively.

We represent the concentration distribution at time $t$ by the column vector $x(t)=(x_1(t),x_2(t),\cdots,x_{N_{S}}(t))^{\top}$. Here, $x_{\alpha}(t)$ represents the concentration of $X_{\alpha}$ at time $t$. We assume that $x_{\alpha}(t)$ is positive for all $\alpha\in\mathcal{S}$. In contrast to the case of the MJP, the distribution $x$ does not necessarily satisfy $\sum_{\alpha\in\mathcal{S}}x_{\alpha}(t)=1$, since the total concentration possibly changes through reactions.

The time evolution of the concentration distribution $x(t)$ is given by the rate equation. Using the stoichiometric matrix instead of the incidence matrix, the rate equation is represented by the continuity equation in Eq.~\eqref{continuity eq}. In this case, the net current $J_e(x,t)$ indicates the net reaction rate of reaction $e$. Since we consider reversible reactions, we assume that the net current $J_e(x,t)$ is given by the difference of the forward and reverse fluxes $J^{\pm}_e(x,t)$ as $J_{e}(x,t)=J^{+}_e(x,t)-J^{-}_e(x,t)$. The fluxes describe the unidirectional reaction rates of the forward and reverse reactions. Here, we also assume that all fluxes are positive $J^{\pm}_e(x,t)>0$. In general, these fluxes $J^{\pm}_e(x,t)$ depend on the concentration distribution. For example, assuming the mass action kinetics, the forward and reverse fluxes are given by $J_{e}^{\pm}(x,t)\coloneqq\kappa_{e}^{\pm}(t)\prod_{\alpha\in\mathcal{S}}[x_{\alpha}(t)]^{n_{\alpha e}^{\pm}}$ with the reaction rate constant $\kappa_{e}^{\pm}(t)$.

\subsection{Thermodynamics}

\subsubsection{Steady state and equilibrium state}
Before considering thermodynamic quantities, we introduce the concept of steady state and equilibrium state. In the following, we use $0$ to indicate a column vector of dimension $N_S$ or $N_E$ with all components equal to zero.  

The system is said to be in a steady state when its time evolution vanishes, i.e., when $d_tx=0$ holds. Due to the continuity equation~\eqref{continuity eq}, the current $J$ satisfies $\nabla^{\top}J=0$ in a steady state. The system is also said to be in an equilibrium state (or simply in equilibrium) when all currents vanish, i.e., when $J=0$ holds. The system is in a steady state if it is in equilibrium, since $J=0$ immediately yields $\nabla^{\top}J=0$. We also refer to a steady state that is not an equilibrium state as a nonequilibrium steady state.

We also introduce the \textit{detailed balance condition}, which guarantees the existence of an equilibrium state. If there exists a distribution $x^{\mathrm{eq}}(t)$ that satisfies $J(x^{\mathrm{eq}}(t),t)=0$, the system is said to satisfy the detailed balance condition at time $t$. Here, $x^{\mathrm{eq}}(t)$ indicates the equilibrium state at time $t$.


\subsubsection{Thermodynamic force and entropy production rate}
For the MJP and the CRN, we can introduce the thermodynamic force and the EPR in the same manner. In the following, we set the Boltzmann constant to one if we consider the MJP, and we take the gas constant as one if we consider the CRN. 

We formally define the thermodynamic force for edge (reaction) $e$ as
\begin{align}
    F_e(x,t)=\ln\frac{J^{+}_e(x,t)}{J^{-}_e(x,t)}.
     \label{thermodynamic force}
\end{align} 
We also simply refer to $F_e$ as the force for edge (reaction) $e$. We also introduce the column vector of the forces as $F(x,t)\coloneqq(F_1(x,t),F_2(x,t),\cdots,F_{N_{E}}(x,t))^{\top}$. As in the case of the currents and fluxes, we will omit the arguments. We assume that the system satisfies the \textit{local detailed balance}~\cite{kondepudi2014modern,beard2007relationship,maes2021local}. In the case of the MJP, this assumption enables us to interpret $F_e$ as the increase of thermodynamic entropy of the system and the thermodynamic reservoirs through the transition on edge $e$. In the case of the CRN, this assumption enables us to interpret $F_e$ as the increase of thermodynamic entropy of the solution and its environment (e.g., particle reservoir of the external species) through reaction $e$.

Using the forces, we define the EPR as
\begin{align}
    \sigma\coloneqq\sum_{e\in\mathcal{E}}J_{e}F_{e}.
    \label{EPR}
\end{align}
We can immediately obtain the second law of thermodynamics $\sigma\geq0$, since the signs of $J_e$ and $F_e$ match for all $e\in\mathcal{E}$. Taking time integral of the EPR leads to the EP in the finite time duration $[0,\tau]$ as
\begin{align}
    \Sigma_{\tau}\coloneqq\int_{0}^{\tau}dt\,\sigma.
    \label{EP}
\end{align}

The definition of $F_e$~\eqref{thermodynamic force} and the relation $J_e=J^{+}_e-J^{-}_e$ allow us to rewrite the EPR in Eq.~\eqref{EPR} as~\cite{yoshimura2021information}
\begin{align}
    \sigma=D_{\mathrm{KL}}(J^{+}\|J^{-})+D_{\mathrm{KL}}(J^{-}\|J^{+}).
    \label{KL form EPR}
\end{align}
Here, the Kullback--Leibler (KL) divergence between two $N_E$-dimensional vectors with positive elements, $K^{(1)}$ and $K^{(2)}$, is defined as
\begin{align}
    D_{\mathrm{KL}}(K^{(1)}\|K^{(2)})\coloneqq\sum_{e\in\mathcal{E}}\left(K^{(1)}_e\ln\frac{K^{(1)}_e}{K^{(2)}_e}-K^{(1)}_e+K^{(2)}_e\right).
\end{align}
The KL divergence $D_{\mathrm{KL}}(K^{(1)}\|K^{(2)})$ becomes zero only when $K^{(1)}=K^{(2)}$ holds. Due to this fact, equation~\eqref{KL form EPR} implies that the EPR becomes zero only when $J^+=J^-$ holds, i.e., the system is in equilibrium. Thus, the EPR measures the irreversibility of the system, that is, the degree of nonequilibrium.

\subsubsection{Conservativeness}
We introduce the \textit{conservativeness} of the forces, which is closely related to the detailed balance condition.
If there exists a potential $\psi(t)=(\psi_{1}(t),\psi_{2}(t),\cdots,\psi_{N_{S}}(t))^{\top}$ that satisfies
\begin{align}
    F(t)=-\nabla\psi(t),
    \label{conservative}
\end{align}
the forces are said to be conservative at time $t$. If we consider the MJP or the CRN with mass action kinetics, we can prove the equivalence of the following two statements: (i) the system satisfies the detailed balance condition at time $t$, and (ii) the forces are conservative at time $t$~\cite{schuster1989generalization} (see Appendix~\ref{ap:conservativeness} for the proof).

\section{Means of forward and reverse fluxes as activities}

In stochastic thermodynamics, some means of forward and reverse fluxes are used to quantify the kinetic activity of MJPs. For example, the arithmetic mean, the geometric mean, and the logarithmic mean are used to define the dynamical activity~\cite{Maes2008,shiraishi2018speed}, the frenetic activity~\cite{maes2017frenetic}, and  the dynamical state mobility~\cite{van2023thermodynamic} (or the edgewise Onsager coefficient~\cite{yoshimura2023housekeeping}), respectively. Even in CRNs, the means of forward and reverse fluxes measure the intensity of reaction. In the following, we let $\mathbb{R}_{\geq0}$ denote the set of all nonnegative real numbers.

\subsection{Homogeneous symmetric mean}
A bivariate function $m:\mathbb{R}_{\geq0}\times\mathbb{R}_{\geq0}\to\mathbb{R}_{\geq0}$ is a \textit{homogeneous symmetric mean} if $m(a,b)$ satisfies the following three properties for all $a,b\geq0$: (i) upper and lower bounded as $\min(a,b)\leq m(a,b)\leq \max(a,b)$, (ii) symmetry ($m(a,b)=m(b,a)$), and (iii) homogeneity ($m(\lambda a,\lambda b)=\lambda m(a,b)$ for all $\lambda\geq0$). We introduce some typical examples of homogeneous symmetric means in TABLE~\ref{tab:means}. All the means that have been used as activities are homogeneous symmetric mean. In the following, we use the term \textit{mean} to refer to homogeneous symmetric mean.

The symmetry and the homogeneity let us characterize $m$ as
\begin{align}
    m(a,b)=bf_m\left(\frac{a}{b}\right),
\end{align}
with the representing function $f_m(r)\coloneqq m(r,1)=m(1,r)\;(r\geq0)$~\cite{kubo1980means,besenyei2012hasegawa}. Reflecting the symmetry and homogeneity of $m$, its representing function $f_m$ satisfies the relation
\begin{align}
    f_m(r)=m(r,1)=rm\left(1,\frac{1}{r}\right)=rf_m\left(\frac{1}{r}\right).
    \label{symmetry_fM}
\end{align}
If $f_m(r)$ is differentiable, we also obtain
\begin{align}
    f_m(r)-rf_m'(r)=f_m'\left(\frac{1}{r}\right),
    \label{symmetry_fM'}
\end{align}
by taking $r$-derivative of the both sides in Eq.~\eqref{symmetry_fM}. \add{Here and below, we let $g'$ and $g''$ denote the first and second derivatives of a function $g$.}

The representing function $f_m(r)$ also characterizes the hierarchy of means. We define a homogeneous symmetric mean $m_1$ to be smaller than another $m_2$ if $m_1(a,b)\leq m_2(a,b)$ holds for all $a,b\geq0$, which we simply write $m_1\leq m_2$. The representing function simplifies this condition into
\begin{align}
    \forall r\geq0,\;f_{m_1}(r)\leq f_{m_2}(r).
    \label{condition hierarchy of means}
\end{align}
We note that the typical means in TABLE~\ref{tab:means} are listed in ascending order in terms of this inequality between means.

Several families of means enable us to treat many means in a \add{single} stroke. For example, the Stolarsky mean~\cite{stolarsky1975generalizations,cisbani1938contributi,tobey1967two} includes the means in TABLE~\ref{tab:means} except the contraharmonic mean. The Stolarsky mean is defined for any $(p,q)\in\mathbb{R}^2$ as
\begin{align}
    S_{p,q}(a,b)\coloneqq
    \begin{cases}
    \left[\dfrac{q(a^p-b^p)}{p(a^q-b^q)}\right]^{\frac{1}{p-q}}& (pq(p-q)\neq0, a\neq b) \\
    \left[\dfrac{a^p-b^p}{p(\ln a-\ln b)}\right]^{\frac{1}{p}}
    & (p\neq0, q=0, a\neq b)\\
    \left[\dfrac{a^q-b^q}{q(\ln a-\ln b)}\right]^{\frac{1}{q}} & (p=0, q\neq 0, a\neq b)\\
    \mathrm{e}^{-\frac{1}{p}}\left(\dfrac{a^{a^p}}{b^{b^p}}\right)^{\frac{1}{a^p-b^p}} & (p=q\neq 0, a\neq b)\\
    \sqrt{ab} & (p=q=0, a\neq b)\\
    a & (a=b)
    \end{cases}
    .
    \label{Stolarsky}
\end{align}
We may regard the Stolarsky mean as one of the most general families of homogeneous symmetric means. This is because the Stolarsky mean includes some well-known families of homogeneous symmetric means parametrized by a real value. For example, taking $p=2q$, we can reduce the Stolarsky mean to the H\"{o}lder mean~\cite{bullen2013handbook} as,
\begin{align}
    S_{2q,q}(a,b)\coloneqq\left(\frac{a^q+b^q}{2}\right)^{\frac{1}{q}}.
    \label{Holder}
\end{align}
This family of means is also called the power mean. The means in TABLE~\ref{tab:means} except the contraharmonic mean and the logarithmic mean are included in the H\"{o}lder mean. Taking $q=1$, we can also reduce the Stolarsky mean to the mean introduced by Galvani~\cite{galvani1927} as
\begin{align}
    S_{p,1}(a,b)\coloneqq
    \begin{cases}
    \left(\dfrac{a^p-b^p}{p(a-b)}\right)^{\frac{1}{p-1}}
    & (a\neq b)\\
    a & (a=b)
    \end{cases}
    .
    \label{Galvani}
\end{align}
Note that some references refer to this family of means as the Stolarsky mean instead of Eq.~\eqref{Stolarsky}. The means in TABLE~\ref{tab:means} except the contraharmonic mean and the harmonic mean are included in this family.


\begin{table}[h]
    \centering
    \caption{Typical homogeneous symmetric means. Here the value of $L(a,a)$ is defined as $a$. The means are listed in \add{ascending order}.}
    \renewcommand{\arraystretch}{2.7}
    \begin{tabular}{cc}
    \hline
    \hspace*{40pt}\raisebox{3pt}{Name}\hspace*{40pt} & \hspace*{30pt}\raisebox{3pt}{Concrete form}\hspace*{30pt}\\
    \hline\hline
    \rowcolor[gray]{.97}[\tabcolsep]
    \raisebox{3pt}{Minimum} & \raisebox{3pt}{$\min(a,b)$}\\
    \raisebox{3pt}{Harmonic mean} & \raisebox{3pt}{$H(a,b)\coloneqq\dfrac{2ab}{a+b}$} \\
    \rowcolor[gray]{.97}[\tabcolsep]
    \raisebox{3pt}{Geometric mean} & \raisebox{3pt}{$G(a,b)\coloneqq\sqrt{ab}$}\\
    \raisebox{3pt}{Logarithmic mean} & \raisebox{3pt}{$L(a,b)\coloneqq\dfrac{a-b}{\ln a-\ln b}$} \\
    \rowcolor[gray]{.97}[\tabcolsep]
    \raisebox{3pt}{Arithmetic mean} & \raisebox{3pt}{$A(a,b)\coloneqq\dfrac{a+b}{2}$}\\
    \raisebox{3pt}{Contraharmonic mean} & \raisebox{3pt}{$C(a,b)\coloneqq\dfrac{a^2+b^2}{a+b}$} \\
    \rowcolor[gray]{.97}[\tabcolsep]
    \raisebox{3pt}{Maximum} & \raisebox{3pt}{$\max(a,b)$} \\
    \hline
    \end{tabular}
    \label{tab:means}
\end{table}

\subsection{General activity}
We define the edgewise activity on edge $e$ with a homogeneous symmetric mean $m$ as
\begin{align}
    \mu_{m,e}\coloneqq m(J^+_e,J^-_e).
\end{align}
In this definition, the average of the forward and reverse fluxes on edge $e$ is measured with the mean $m$. We also define the activity measured with $m$ as
\begin{align}
    \mu_m\coloneqq\sum_{e\in \mathcal{E}}\mu_{m,e}.
    \label{activity}
\end{align}
This general activity turns into the well-known special cases as follows: the dynamical activity $\sum_{e\in\mathcal{E}}(J^{+}_e+J^{-}_e)$ is given by $2\mu_{A}$, and the dynamical state mobility $\sum_{e\in\mathcal{E}}(J^{+}_e-J^{-}_e)/(\ln J^{+}_e-\ln J^{-}_e)$ is given by $\mu_L$. Here, $A(a,b)$ and $L(a,b)$ corresponding to $\mu_A$ and $\mu_L$ are the arithmetic mean and the logarithmic mean introduced in TABLE~\ref{tab:means}, respectively.

We can regard the general activity~\eqref{activity} as the total intensity of all jumps or reactions. Using different $m$ corresponds to changing how to take the average of the forward and reverse rates. If we consider the continuum limit of MJPs, the general activity corresponds to the diffusion coefficient regardless of the choice of $m$ (see Appendix~\ref{ap:continuum} for details).

If the system is in a steady state, we can relate the activity to the time scale. In this case, the dynamical activity $2\mu_A$ indicates the total number of jumps or reactions per unit time in the steady state. Thus, we can define the time required for a single jump or reaction to occur as $\mathcal{T}_0\coloneqq(2\mu_A)^{-1}$. If we use $m$ such that $m \leq A$, the general activity $\mu_m$ provides an upper bound of $\mathcal{T}_0$ as $ \mathcal{T}_0\leq(2\mu_m)^{-1}$. Conversely, if we use $m \geq A $, $\mu_m$ provides a lower bound as $ \mathcal{T}_0\geq(2\mu_m)^{-1}$.

\subsection{Physical conditions on means}
In the following, we only use homogeneous symmetric means whose representing function $f_m(r)$ is second-order differentiable at $r>1$. We also impose the following two conditions on $f_m(r)$:
\begin{align}
    \forall r>1,\;f_m'(r)+f_m'\left(\frac{1}{r}\right)>0,
    \label{condition_fM'}
\end{align}
and
\begin{align}
    \forall r>1,\;f_m''(r)\geq-\frac{r+1}{r(r-1)^2}\left[f_m'(r)+f_m'\left(\frac{1}{r}\right)\right].
    \label{condition_fM''}
\end{align}
These two conditions are essential to obtain TSLs. While seemingly complex, these conditions are physically meaningful. We reveal the physical interpretation of each condition in the following two sections and Appendix~\ref{ap:rewriting conditions}. Furthermore, these conditions are satisfied by broad means. For example, we can verify that the Stolarsky mean~\eqref{Stolarsky} satisfies these conditions (see Appendix~\ref{ap:stolarsky} for the proof). We can also verify that the contraharmonic mean, not included in the Stolarsky mean, satisfies the conditions (see Appendix~\ref{ap:contraharmonic} for the proof). Thus, all of the typical means in TABLE~\ref{tab:means} satisfy the conditions.

\subsubsection{One by one relation between current and force induced by general activity}
\label{sec:force current relation}
To reveal the meaning of the first condition~\eqref{condition_fM'}, we introduce the fact that the activity enables us to relate the current and force. We can express $J_e$ as a function of $F_e$ with the activity $\mu_{m,e}$ as
\begin{align}
    J_{e}=\mu_{m,e}\Psi_{m}(F_{e}),
    \label{relation FtoJ}
\end{align}
where $\Psi_{m}(u)$ is defined as 
\begin{align}
    \Psi_{m}(u)\coloneqq\frac{\mathrm{e}^u-1}{f_{m}(\mathrm{e}^u)}.
    \label{Psi_m}
\end{align}
This relation~\eqref{relation FtoJ} can be regarded as a nonlinear extension of the Onsager reciprocal relation~\cite{onsager1931reciprocal,casimir1945onsager,mielke2016generalization,vroylandt2018degree,yoshimura2023housekeeping,kobayashi2023information}. It is derived as follows: Using $J^+_e/J^-_e=\mathrm{e}^{F_e}$, we can represent $J_e$ and $\mu_{m,e}$ with $J^{-}_{e}$ and $F_e$ as $J_{e}=J^{-}_{e}(\mathrm{e}^{F_e}-1)$ and $\mu_{m,e}=J^{-}_{e}f_{m}(J^+_e/J^-_e)=J^{-}_{e}f_{m}(\mathrm{e}^{F_{e}})$. Eliminating $J^-_e$ from these two equations and using the definition of $\Psi_m$~\eqref{Psi_m}, we obtain the relation~\eqref{relation FtoJ}. We remark that the function $\Psi_{m}$ is odd, which is verified by using the relation~\eqref{symmetry_fM} as
\begin{align}
    \Psi_{m}(-u)=\frac{\mathrm{e}^{-u}-1}{f_{m}(\mathrm{e}^{-u})}=\frac{\mathrm{e}^{-u}-1}{\mathrm{e}^{-u}f_{m}(\mathrm{e}^{u})}=-\Psi_{m}(u).
\end{align}
Due to this oddness, the relation~\eqref{relation FtoJ} connects $-F_e$ to $-J_e$ when it connects $F_e$ to $J_e$. In particular, the relation links the force at the equilibrium $F_e=0$ to the current at the equilibrium $J_e=0$.

Physically, the condition~\eqref{condition_fM'} yields the inverse version of Eq.~\eqref{relation FtoJ}: we can express $F_e$ as a function of $J_e$ as
\begin{align}
    F_{e}=\Psi_{m}^{-1}\left(\frac{J_{e}}{\mu_{m,e}}\right).
    \label{relation JtoF}
\end{align}
Here, the existence of the inverse function $\Psi_m^{-1}$ is verified by the fact that the condition~\eqref{condition_fM'} is equivalent to the monotonically increasingness of $\Psi_m$ (see also Appendix~\ref{ap:rewriting first condition} for details). The inverse function $\Psi_{m}^{-1}$ is odd and monotonically increasing since $\Psi_m$ is so. In particular, due to the oddness, $-J_e$ is mapped to $-F_e$ in relation~\eqref{relation FtoJ} when $J_e$ mapped to $F_e$.

We now clarify the domain of the function $\Psi_m^{-1}$. While $\Psi_m$ is defined over the entire set of real numbers, the domain of $\Psi_m^{-1}$ is restricted to the range of $\Psi_m$. Specifically, $\Psi_m^{-1}(\omega)$ is defined only for $\omega$ satisfying $-\Psi_m(\infty)\leq\omega\leq\Psi_m(\infty)$ because $\Psi_m(u)$ is a monotonically increasing function and $\Psi_m(-u)=-\Psi_m(u)$.  Here, we let $\Psi_m(\infty)$ denote the (possibly infinite) limit $\lim_{u\to\infty}\Psi_m(u)$. We note that Eq.~\eqref{relation FtoJ} allows $J_e/\mu_{m,e}$ be included in the domain of $\Psi_m^{-1}$ for all $e\in\mathcal{E}$.

We remark that the relations in Eq.~\eqref{relation FtoJ} and Eq.~\eqref{relation JtoF} let the current and the force be mutually conjugate variables in terms of the Legendre duality. This duality is investigated and used to study the gradient structure of MJPs and CRNs~\cite{mielke2011gradient,maas2011gradient,chow2012fokker,mielke2013geodesic, kaiser2018canonical,yoshimura2023housekeeping, van2023thermodynamic,mielke2014relation,mielke2017non,kaiser2018canonical,maes2017frenetic,kobayashi2022hessian,peletier2022jump} (see also Appendix~\ref{ap:duality} for details). In particular, the duality with a general homogeneous symmetric mean has been discussed in Ref.~\cite{peletier2022jump}. However, they mainly focus on means whose representing function is concave for their mathematical purpose.

\subsubsection{Monotonicity of entropy production rate}
\label{sec:EPR convexity}

To reveal the meaning of the second condition~\eqref{condition_fM''}, we rewrite the EPR using the relation in Eq.~\eqref{relation JtoF} as 
\begin{align}
    \sigma&=\sum_{e\in \mathcal{E}}J_e\Psi_{m}^{-1}\left(\frac{J_{e}}{\mu_{m,e}}\right).
    \label{rewrite EPR}
\end{align}
Since the right-hand side can be rewritten as $\sum_{e\in \mathcal{E}}\mu_{m,e}(J_{e}/\mu_{m,e})\Psi_{m}^{-1}(J_{e}/\mu_{m,e})$, the new representation~\eqref{rewrite EPR} is characterized by a function $w\Psi_m^{-1}(w)$. We note that this function is even and monotonically increasing on $w>0$ due to the oddness and monotonicity of $\Psi_m^{-1}$.

Under the first condition~\eqref{condition_fM'}, the second condition~\eqref{condition_fM''} is equivalent to the convexity of $w\Psi_m^{-1}(w)$ (see also Appendix~\ref{ap:rewriting second condition} for details). This convexity implies that the EPR monotonically decreases by coarse-graining two edges $e_1$ and $e_2$ into one edge $e_{0}$ in the following way: The quantity on the coarse-grained edge is given by the sum of the quantities on the original edges as $J_{e_{0}}=J_{e_1}+J_{e_2}$ and $\mu_{m,e_{0}}=\mu_{m,e_1}+\mu_{m,e_2}$. Indeed, this monotonicity of the EPR is obtained by the convexity of $w\Psi_m^{-1}(w)$ as follows:
\begin{align*}
    &J_{e_1}\Psi_{m}^{-1}\left(\frac{J_{e_1}}{\mu_{m,e_1}}\right)+J_{e_2}\Psi_{m}^{-1}\left(\frac{J_{e_2}}{\mu_{m,e_2}}\right)\\
    &=\mu_{m,e_{0}}\sum_{i=1,2}\frac{\mu_{m,e_i}}{\mu_{m,e_{0}}}\left\{\frac{J_{e_i}}{\mu_{m,e_i}}\Psi_{m}^{-1}\left(\frac{J_{e_i}}{\mu_{m,e_i}}\right)\right\}\\
    &\geq J_{e_{0}}\Psi_{m}^{-1}\left(\frac{J_{e_{0}}}{\mu_{m,e_{0}}}\right).
\end{align*}
Here, $J_{e_{0}}/\mu_{m,e_{0}}$ is included in the domain of $\Psi_m^{-1}$, since this quantity is the convex combination of $J_{e_{1}}/\mu_{m,e_{1}}$ and $J_{e_{2}}/\mu_{m,e_{2}}$.

\section{Thermodynamic speed limits with the 1-Wasserstein distance}

\subsection{Bounds for general currents}

We here introduce a trade-off relation which is essential for the derivation of the TSLs. The trade-off relation provides a lower bound of the time average of the EPR  $\langle\sigma\rangle_{\tau}$ by the speed of the time evolution of a general observable.

The monotonically decreasingness of the EPR discussed in Sec.~\ref{sec:EPR convexity} yields the trade-off relation between the dissipation and the intensity of a general current. The trade-off relation is represented as the following lower bound of $\langle\sigma\rangle_{\tau}$ with a general current $\mathcal{J}_{c}(t)\coloneqq\sum_{e\in \mathcal{E}}c_e(t)J_e(t)$:
\begin{align}
    \langle\sigma\rangle_{\tau}\geq\frac{\langle|\mathcal{J}_{c}|\rangle_{\tau}}{|c|_{\infty}}\Psi_m^{-1}\left(\frac{\langle|\mathcal{J}_{c}|\rangle_{\tau}}{|c|_{\infty}\langle\mu_m\rangle_{\tau}}\right).
    \label{finitetimeTUR}
\end{align}
Here, we let $|c|_{\infty}$ denote $\max_{e\in \mathcal{E},t\in[0,\tau]}|c_e(t)|$. We also define the time average as $\langle\bullet\rangle_{\tau}=(1/\tau)\int_{0}^{\tau}dt\,\bullet$. The lower bound in Eq.~\eqref{finitetimeTUR} is monotonically increasing with $\langle|\mathcal{J}_{c}|\rangle_{\tau}$ and monotonically decreasing with $\langle\mu_m\rangle_{\tau}$. Thus, this bound implies that a larger EP is required to realize a more intense current with lower activity. 

This inequality~\eqref{finitetimeTUR} is derived by applying Jensen's inequality for the convex function $\omega\Psi_m^{-1}(\omega)$ (see Appendix~\ref{ap:generalcurrents} for the proof). In Appendix~\ref{ap:statewisebound}, we also discuss an application of the trade-off relation as a bound for statewise observables.

\begin{table*}
    \centering
    \caption{The TSLs with typical means. We also show the concrete form of the edgewise activities and the functions $\Psi_m(u)$ and $\Psi_m^{-1}(\omega)$. Due to Eq.~\eqref{speed in domain}, we only show the forms of $\Psi_m(u)$ and $\Psi_m^{-1}(\omega)$ on $u\geq0$ and $0\leq\omega\leq\Psi_m(\infty)$, respectively}
    \renewcommand{\arraystretch}{2.7}
    \begin{tabular}{ccccc}
    \hline
    \hspace*{40pt}\raisebox{3pt}{Mean\,($m$)}\hspace*{40pt} & \hspace*{30pt}\raisebox{3pt}{$\mu_{m,e}$}\hspace*{30pt} & \hspace*{10pt}\raisebox{3pt}{\renewcommand{\arraystretch}{1.5}\begin{tabular}{c}$\Psi_m(u)$\\$(u\geq0)$\end{tabular}}\hspace*{10pt} &\hspace*{20pt}\raisebox{3pt}{\renewcommand{\arraystretch}{1.5}\begin{tabular}{c}$\Psi_m^{-1}(\omega)$\\$(0\leq\omega\leq\Psi_m(\infty))$\end{tabular}}\hspace*{20pt} & \hspace*{40pt}\raisebox{3pt}{TSL}\hspace*{40pt}\\
    \hline\hline
    \rowcolor[gray]{.97}[\tabcolsep]
    \raisebox{3pt}{Minimum\,($\min$)} & \raisebox{3pt}{$\min(J^+_e,J^-_e)$} & \raisebox{3pt}{$\mathrm{e}^u-1$} &\raisebox{3pt}{$\ln(1+\omega)$} & \raisebox{3pt}{$\langle\sigma\rangle_{\tau}\geq\langle v_1\rangle_{\tau}\ln\left(1+\dfrac{\langle v_1\rangle_{\tau}}{\langle \mu_m\rangle_{\tau}}\right)$}\\
    \raisebox{3pt}{Harmonic mean\,($H$)} & \raisebox{3pt}{$\dfrac{2J^+_eJ^-_e}{J^+_e+J^-_e}$} & \raisebox{3pt}{$\sinh u$} & \raisebox{3pt}{$\sinh^{-1} \omega$} & \raisebox{3pt}{$\langle\sigma\rangle_{\tau}\geq\langle v_1\rangle_{\tau}\sinh^{-1}\left(\dfrac{\langle v_1\rangle_{\tau}}{\langle \mu_m\rangle_{\tau}}\right)$} \\
    \rowcolor[gray]{.97}[\tabcolsep]
    \raisebox{3pt}{Geometric mean\,($G$)} & \raisebox{3pt}{$\sqrt{J^+_eJ^-_e}$} & \raisebox{3pt}{$2\sinh\dfrac{u}{2}$} & \raisebox{3pt}{$2\sinh^{-1}\dfrac{\omega}{2}$} & \raisebox{3pt}{$\langle\sigma\rangle_{\tau}\geq2\langle v_1\rangle_{\tau}\sinh^{-1}\left(\dfrac{\langle v_1\rangle_{\tau}}{2\langle \mu_m\rangle_{\tau}}\right)$} \\
    \raisebox{3pt}{Logarithmic mean\,($L$)} & \raisebox{3pt}{$\dfrac{J^+_e-J^-_e}{\ln J^+_e-\ln J^-_e}$} & \raisebox{3pt}{$u$} & \raisebox{3pt}{$\omega$} & \raisebox{3pt}{$\langle\sigma\rangle_{\tau}\geq\dfrac{\langle v_1\rangle_{\tau}^2}{\langle \mu_m\rangle_{\tau}}$} \\
    \rowcolor[gray]{.97}[\tabcolsep]
    \raisebox{3pt}{Arithmetic mean\,($A$)} & \raisebox{3pt}{$\dfrac{J^+_e+J^-_e}{2}$}& \raisebox{3pt}{$2\tanh\dfrac{u}{2}$} & \raisebox{3pt}{$2\tanh^{-1}\dfrac{\omega}{2}$} & \raisebox{3pt}{$\langle\sigma\rangle_{\tau}\geq2\langle v_1\rangle_{\tau}\tanh^{-1}\left(\dfrac{\langle v_1\rangle_{\tau}}{2\langle \mu_m\rangle_{\tau}}\right)$} \\
    \raisebox{3pt}{Contraharmonic mean\,($C$)} & \raisebox{3pt}{$\dfrac{(J^{+}_e)^2+(J^{-}_e)^2}{J^+_e+J^-_e}$} & \raisebox{3pt}{$\tanh u$} & \raisebox{3pt}{$\tanh^{-1} \omega$} & \raisebox{3pt}{$\langle\sigma\rangle_{\tau}\geq\langle v_1\rangle_{\tau}\tanh^{-1}\left(\dfrac{\langle v_1\rangle_{\tau}}{\langle \mu_m\rangle_{\tau}}\right)$}\\
    \rowcolor[gray]{.97}[\tabcolsep]
    \raisebox{3pt}{Maximum\,($\max$)} & \raisebox{3pt}{$\max(J^+_e,J^-_e)$} & \raisebox{3pt}{$1-\mathrm{e}^{-u}$} & \raisebox{3pt}{$-\ln(1-\omega)$} & \raisebox{3pt}{$\langle\sigma\rangle_{\tau}\geq-\langle v_1\rangle_{\tau}\ln\left(1-\dfrac{\langle v_1\rangle_{\tau}}{\langle \mu_m\rangle_{\tau}}\right)$}\\
    \hline
    \end{tabular}
    \label{tab:TSLs}
\end{table*}

\subsection{1-Wasserstein distance and speed of the time evolution}
We define the 1-Wasserstein distance between the two distributions $x^A$ and $x^B$ via the following minimization problem,
\begin{align}
    W_1(x^A,x^B)=\inf_{U}\sum_{e\in\mathcal{E}}\left|U_e\right|,
    \label{Beckmann problem}
\end{align}
where the infimum is over all $U=(U_1,U_2,\cdots,U_{N_{E}})^{\top}$ satisfying $x^B-x^A=\nabla^{\top}U$. In the case of the MJPs, it is well known as the Beckmann problem~\cite{beckmann1952continuous,peyre2019computational}, which is generalized to chemical systems in nonequilibrium thermodynamics~\cite{nagayama2023geometric}. Note that the condition on $U$ lets us define the 1-Wasserstein distance between $x^A$ and $x^B$ only when $x^B-x^A$ belongs to the image of $\nabla^{\top}$.

We remark that the Beckmann problem is the dual problem of another optimization problem, which is called the Kantorovich--Rubinstein duality~\cite{villani2009optimal,peyre2019computational}. We also introduce this duality formula in Appendix~\ref{ap:KR dual}.


The 1-Wasserstein distance enables us to measure the speed of the time evolution as
\begin{align}
    v_1(t)\coloneqq\lim_{\varDelta t\to0}\frac{W_1(x(t),x(t+\varDelta t))}{\varDelta t}.
    \label{Wasserstein speed}
\end{align}
The Beckmann problem~\eqref{Beckmann problem} provides the upper bound of the speed,
\begin{align}
    v_1(t)\leq\sum_{e\in\mathcal{E}}|J_e(t)|.
    \label{speed_ineq}
\end{align}
This bound is obtained as follows. Since $x(t+\varDelta t)-x(t)=\varDelta t\nabla^{\top}J(t)+O(\varDelta t^2)$ holds, $\varDelta tJ(t)+O(\varDelta t^2)$ is a candidate of the optimization problem corresponding to $W_1(x(t),x(t+\varDelta t))$. It immediately leads to $W_1(x(t),x(t+\varDelta t))\leq\varDelta t\sum_{e\in\mathcal{E}}|J_e(t)|+O(\varDelta t^2)$. Deviding both sides of this inequality by $\varDelta t$ and taking the limit $\varDelta t\to0$ conclude the desired bound~\eqref{speed_ineq}.

\subsection{Thermodynamic speed limits with the 1-Wasserstein distance}

Using the speed measured with the 1-Wasserstein distance~\eqref{Wasserstein speed} and the general activity, we can obtain a series of TSLs, 
\begin{align}
    \langle\sigma\rangle_{\tau}&\geq \langle v_1\rangle_{\tau}\Psi_m^{-1}\left(\frac{\langle v_1\rangle_{\tau}}{\langle\mu_m\rangle_{\tau}}\right)\label{TSLs}\\
    &\geq\frac{W_1(x(0),x(\tau))}{\tau}\Psi_m^{-1}\left(\frac{W_1(x(0),x(\tau))}{\tau\langle\mu_m\rangle_{\tau}}\right).
    \label{TSLs2}
\end{align}
The equation~\eqref{TSLs} is derived from the lower bound of the EP~\eqref{finitetimeTUR} and the inequality for the speed~\eqref{speed_ineq} as follows. Taking the coefficient $c_e$ in the trade-off relation~\eqref{finitetimeTUR} as
\begin{align}
    c_e(t)\coloneqq\begin{cases}
        1 & (J_e(t)\geq0)\\
        -1 & (J_e(t)<0)
    \end{cases},
\end{align}
we obtain
\begin{align}
    \langle\sigma\rangle_{\tau}\geq\left\langle\sum_{e\in\mathcal{E}}|J_e|\right\rangle_{\tau}\Psi_m^{-1}\left(\frac{\left\langle\sum_{e\in\mathcal{E}}|J_e|\right\rangle_{\tau}}{\langle\mu_m\rangle_{\tau}}\right),
\end{align}
since $|c|_{\infty}=1$. Combining this inequality and Eq.~\eqref{speed_ineq}, we can obtain Eq.~\eqref{TSLs} because the function $\omega\Psi_m^{-1}(\omega)$ is monotonically increasing on $\omega\geq0$. The inequality in Eq.~\eqref{TSLs2} is also obtained by the triangle inequality of the 1-Wasserstein distance $\int_{0}^{\tau}dt\,v_1(t)\geq W_1(x(0),x(\tau))$. We also provide another proof with the Kantorovich--Rubinstein duality in Appendix~\ref{ap:anotherproofTSL}.

We note that the arguments of $\Psi_m^{-1}$ appearing in Eqs.~\eqref{TSLs} and~\eqref{TSLs2} are included in the nonnegative domain $0\leq\omega\leq\Psi_m(\infty)$. This is because the following inequalities hold:
\begin{align}
    0\leq\frac{W_1(x(0),x(\tau))}{\tau\langle\mu_m\rangle_{\tau}}\leq\frac{\langle v_1\rangle_{\tau}}{\langle\mu_m\rangle_{\tau}}\leq\Psi_m(\infty).
    \label{speed in domain}
\end{align}
Here, the first and the second inequalities follow from the nonnegativity and the triangle inequality of the 1-Wasserstein distance, respectively. The last one is obtained as 
\begin{align}
    \frac{\langle v_1\rangle_{\tau}}{\langle\mu_m\rangle_{\tau}}\leq\frac{\int_{0}^{\tau}dt\sum_{e\in\mathcal{E}} |J_e|}{\int_{0}^{\tau}dt\sum_{e\in\mathcal{E}}\mu_{m,e}}\leq\max_{e\in\mathcal{E},t\in[0,\tau]}\frac{|J_e|}{\mu_{m,e}}\leq\Psi_m(\infty),
\end{align}
where we use Eq.~\eqref{relation FtoJ}, Eq.~\eqref{speed_ineq}, the monotonically increasingness of $\Psi_m$, and the following inequality: 
\begin{align}
&\int_{0}^{\tau}dt\sum_{e\in\mathcal{E}} \mu_{m,e}\frac{|J_e|}{\mu_{m,e}}\notag\\&\leq \left( \int_{0}^{\tau}dt\sum_{e\in\mathcal{E}}  \mu_{m,e} \right)\left( \max_{e\in\mathcal{E},t\in[0,\tau]}\frac{|J_e|}{\mu_{m,e}} \right).
\end{align}

Physically, the TSL~\eqref{TSLs} represent the trade-off relation between three elements, the dissipation, the speed of the time evolution, and the activity. Since the lower bound $\langle v_1\rangle_{\tau}\Psi_m^{-1}(\langle v_1\rangle_{\tau}/\langle\mu_m\rangle_{\tau})$ increases with respect to $\langle v_1\rangle_{\tau}$ and decreases with respect to $\langle\mu_m\rangle_{\tau}$, we need greater dissipation to evolve the system with faster speed or smaller activity.

Substituting various means satisfying the conditions in Eqs.~\eqref{condition_fM'} and~\eqref{condition_fM''} into the general forms in Eqs.~\eqref{TSLs} and~\eqref{TSLs2}, we can obtain an infinite variety of TSLs. For example, using the Stolarsky mean $S_{p,q}$ as $m$ in Eqs.~\eqref{TSLs} and~\eqref{TSLs2} provides TSLs for all pairs of real numbers $(p,q)$. In general, we need to compute the inverse function $\Psi_m^{-1}$ numerically to verify the TSLs, since it does not have closed forms. However, the TSLs reduce to some simple forms for several means: All typical means in TABLE~\ref{tab:means} let the TSLs be simple as shown in TABLE~\ref{tab:TSLs}. We remark that the TSL with the arithmetic mean
\begin{align}
    \langle\sigma\rangle_{\tau}\geq\frac{2W_1(x(0),x(\tau))}{\tau}\tanh^{-1}\left(\frac{W_1(x(0),x(\tau))}{2\tau\langle\mu_A\rangle_{\tau}}\right),
    \label{TSL with A}
\end{align}
and the one with the logarithmic mean
\begin{align}
    \langle\sigma\rangle_{\tau}\geq\frac{W_1(x(0),x(\tau))^2}{\tau^2\langle\mu_L\rangle_{\tau}},
\end{align}
are the same as the ones in the previous studies~\cite{dechant2022minimum,van2023thermodynamic}.

We can also rewrite the TSLs ~\eqref{TSLs} and~\eqref{TSLs2} into lower bounds of the transition time $\tau$. To do so, we define the path length of the time series of $x$ as $l_{1,\tau}\coloneqq\int_{0}^{\tau}dt\,v_1$. Using the relations $\tau\langle v_1\rangle_{\tau}=l_{1,\tau}$ and $\tau\langle \sigma\rangle_{\tau}=\Sigma_{\tau}$, we obtain
\begin{align}
    \tau&\geq\left[\frac{\langle\mu_m\rangle_{\tau}}{l_{1,\tau}}\Psi_m\left(\frac{\Sigma_{\tau}}{l_{1,\tau}}\right)\right]^{-1}\notag\\
    &\geq\left[\frac{\langle\mu_m\rangle_{\tau}}{W_1(x(0),x(\tau))}\Psi_m\left(\frac{\Sigma_{\tau}}{W_1(x(0),x(\tau))}\right)\right]^{-1}.
    \label{time bounds}
\end{align}
We provide the derivation in Appendix~\ref{ap:time bound}. We can regard the first lower bound in Eq.~\eqref{time bounds} as the minimum time required to evolve the system along the original path $\{x(t)\}_{t\in[0,\tau]}$. The second lower bound in Eq.~\eqref{time bounds} also provides the minimum time required to evolve the system from $x(0)$ to $x(\tau)$, where the path may be different from the original one. In both cases, we can reduce the minimum time by spending more dissipation or achieving greater activity.
In contrast to the original form~\eqref{TSLs} and~\eqref{TSLs2}, we can obtain these lower bounds without calculating the inverse function $\Psi_m^{-1}$.

Historically, TSLs for MJPs are derived using a simpler distance, called the total variation distance, $d_{\mathrm{TV}}(x^A,x^B)\coloneqq\sum_{\alpha\in\mathcal{S}}|x^B_{\alpha}-x^A_{\alpha}|/2$~\cite{shiraishi2018speed,lee2022speed}. This definition is applicable even in the case of CRNs. We can also verify that the total variation distance provides a lower bound of the 1-Wasserstein distance. Thus, the total variation distance leads to weaker TSLs (see also Appendix~\ref{ap:tvd} for details).

\begin{figure*}
    \centering
    \includegraphics[width=\linewidth]{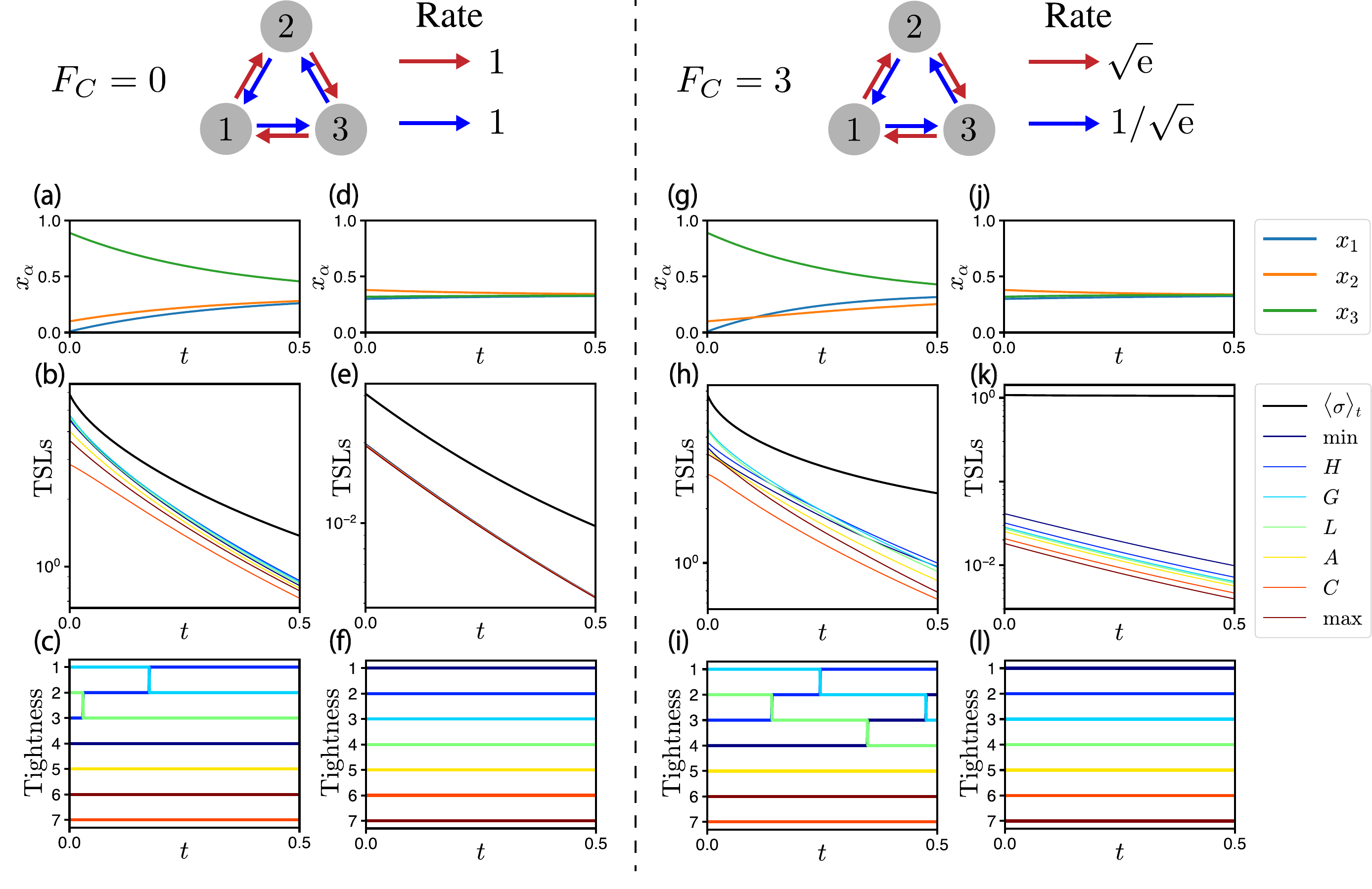}
    \caption{The hierarchy of TSLs~\eqref{TSLs} in TABLE~\ref{tab:TSLs}. We compare the TSLs~\eqref{TSLs} in detailed balanced and driven systems near and far from steady state. The region to the left of the dashed line corresponds to the detailed balanced system ($F_C=0$), while the region to the right corresponds to the driven system ($F_C=3$). We also show the system used in this numerical calculation. The gray circles correspond to microstates. We take the transition rates for moving between microstates clockwise (red arrows) and counterclockwise (blue arrows) as $\mathrm{e}^{F_C/6}$ and $\mathrm{e}^{-F_C/6}$, respectively. (a) Time series of the probability distribution in case (I) ($F_C=0$, $x(0)=(0.01,0.1,0.89)^{\top}$). (b) The TSLs in case (I). (c) The tightness of the TSLs in case (I).
    (d) Time series of the probability distribution in case (II) ($F_C=0$, $x(0)=(0.3,0.38,0.32)^{\top}$). (e) The TSLs in case (II). Since the system is near equilibrium, all the TSLs provide almost the same lower bounds. (f) The tightness of the TSLs in case (II). The TSL becomes tighter with smaller means because the speed of the time evolution is slow. (g) Time series of the probability distribution in case (III) ($F_C=3$, $x(0)=(0.01,0.1,0.89)^{\top}$). (h) The TSLs in case (III). (i) The tightness of the TSLs in case (III). (j) Time series of the probability distribution in case (IV) ($F_C=3$, $x(0)=(0.3,0.38,0.32)^{\top}$). (k) The TSLs in case (IV). (l) The tightness of the TSLs in case (IV). Since the system is near the nonequilibrium steady state, the TSL with a smaller mean provides a tighter bound. In (b), (e), (h), and (k), the black line indicates $\langle\sigma\rangle_{\tau}$. The other lines indicate the lower bounds of $\langle\sigma\rangle_{\tau}$ provided by the TSLs in TABLE~\ref{tab:TSLs} (in the legend, we only indicate the means used as the activity). In (c), (f), (i), and (l), the smaller the number, the tighter the lower bound on the EPR. The colors of lines correspond to the mean used in the TSLs.}
    \label{fig:TL}
\end{figure*}


\subsection{Appearance of hierarchy in low-speed regimes}


In general, a hierarchy of means $m_1 \leq m_2$ does not provide a hierarchy of the TSLs, that is an apparent inequality between the two lower bounds $\langle v_1\rangle_{\tau} \Psi_{m_1}^{-1} (\langle v_1\rangle_{\tau}/\langle\mu_{m_1}\rangle_{\tau})$ and $\langle v_1\rangle_{\tau} \Psi_{m_2}^{-1} (\langle v_1\rangle_{\tau}/\langle\mu_{m_2}\rangle_{\tau})$. Therefore, there is no definite choice of the mean, which provides the tightest bound in the TSLs. In general, we can only obtain the two inequalities $\langle\mu_{m_1}\rangle_{\tau} \leq \langle\mu_{m_2}\rangle_{\tau}$ and
\begin{align}
    \Psi_{m_1}^{-1}(\omega)\leq\Psi_{m_2}^{-1}(\omega),
    \label{ineq_btw_Psi_m}
\end{align} 
for any $0\leq\omega\leq\Psi_{m_2}(\infty)$ when $m_1 \leq m_2$. This is because $\mu_m$ is monotonically increasing in $m$, and Eq.~\eqref{ineq_btw_Psi_m} follows from the fact $\Psi_{m_1}(u)\geq\Psi_{m_2}(u)$ and the monotonically increasingness of $\Psi_{m}$. Here, $\Psi_{m_1}(u)\geq\Psi_{m_2}(u)$ is obtained from $f_{m_1}(r)\leq f_{m_2}(r)$. These two general inequalities do not provide an apparent inequality between the two lower bounds. In fact, we have found numerically that the tightness of the TSLs between two means can change over time as discussed later.

However, it may be possible to show the existence of a hierarchy between TSLs with two specific means. For example, we can show that the TSL with $\min(a,b)$ is always tighter than (or equivalent to) the TSL with $\max(a,b)$ as follows. Combining the relation $\mu_{\max,e}-\mu_{\min,e}=|J_e|$ and the inequality in Eq.~\eqref{speed_ineq}, we obtain
\begin{align}
    \langle\mu_{\max}\rangle_{\tau}-\langle\mu_{\min}\rangle_{\tau}=\left\langle\sum_{e\in\mathcal{E}}|J_e|\right\rangle_{\tau}\geq\langle v_1\rangle_{\tau}.
    \label{diffmaxmin ineq}
\end{align}
Using this inequality~\eqref{diffmaxmin ineq} and nonnegativities of $\langle\mu_{\min}\rangle_{\tau}$ and $\langle v_1\rangle_{\tau}$, we can easily verify that the following inequality holds:
\begin{align}
    \left(1-\frac{\langle v_1\rangle_{\tau}}{\langle\mu_{\max}\rangle_{\tau}}\right)^{-1}\leq 1+\frac{\langle v_1\rangle_{\tau}}{\langle\mu_{\min}\rangle_{\tau}}.
    \label{arguments ineq}
\end{align}
By taking the logarithm of both sides of this inequality~\eqref{arguments ineq} and then multiplying both sides by $\langle v_1\rangle_{\tau}$, we obtain the hierarchy between the TSLs with $\max(a,b)$ and $\min(a,b)$ as
\begin{align}
    -\langle v_1\rangle_{\tau}\ln\left(1-\frac{\langle v_1\rangle_{\tau}}{\langle\mu_{\max}\rangle_{\tau}}\right)\leq\langle v_1\rangle_{\tau}\ln\left(1+\frac{\langle v_1\rangle_{\tau}}{\langle\mu_{\min}\rangle_{\tau}}\right),
    \label{hierarchy minmax}
\end{align}
or equivalently
\begin{align}
\langle v_1\rangle_{\tau} \Psi_{\rm max}^{-1} \left( \frac{\langle v_1\rangle_{\tau}}{\langle\mu_{\rm max}\rangle_{\tau}} \right) \leq \langle v_1\rangle_{\tau} \Psi_{\rm min}^{-1} \left( \frac{\langle v_1\rangle_{\tau}}{\langle\mu_{\rm min}\rangle_{\tau}} \right).
\end{align}

Moreover, we have a hierarchy of the TSLs if the state is close to a steady state: the smaller the mean, the tighter the bound. Close to the steady state, the speed decreases, while the activity remains finite. Thus, we can assume $\langle v_1\rangle_{\tau}\ll\langle\mu_m\rangle_{\tau}$. This assumption allows us to expand $\Psi_m^{-1}(\langle v_1\rangle_{\tau}/\langle\mu_m\rangle_{\tau})$ in the TSL~\eqref{TSLs} using a Taylor series as
\begin{align}
    \Psi_m^{-1}\left(\frac{\langle v_1\rangle_{\tau}}{\langle\mu_m\rangle_{\tau}}\right)= \frac{\langle v_1\rangle_{\tau}}{\langle\mu_m\rangle_{\tau}}+ O\left(\left(\frac{\langle v_1\rangle_{\tau}}{\langle\mu_m\rangle_{\tau}}\right)^3\right).
\end{align}
Here, we used $\Psi_m^{-1}(0)=0$ and $(\Psi_m^{-1})'(0)=1/\Psi_m'(0)=1$. We note that the even-order terms of $\langle v_1\rangle_{\tau}/\langle\mu_m\rangle_{\tau}$ vanish because $\Psi_m^{-1}(\omega)$ is an odd function. Due to this expansion, the TSL reduces to
\begin{align}
    \langle\sigma\rangle_{\tau}\geq\frac{\langle v_1\rangle_{\tau}^2}{\langle\mu_m\rangle_{\tau}}.
    \label{TSL slow}
\end{align}
This form implies that the TSL with a smaller mean becomes tighter.

In addition, the TSL~\eqref{TSLs} gives almost the same bound regardless of the mean used if the system is near equilibrium. In this situation, we can take a small constant $\epsilon_{\mathrm{eq}}$ satisfying $|J_e|\leq\epsilon_{\mathrm{eq}}$ for all $e\in\mathcal{E}$. This is because all currents vanish in equilibrium. Then, we can use the representation in Eq.~\eqref{TSL slow}, since the inequality for $v_1$~\eqref{speed_ineq} yields $v_1\leq\sum_{e\in\mathcal{E}}|J_e|\leq N_{E}\epsilon_{\mathrm{eq}}$. We can also obtain
\begin{align}
    \mu_{m,e}=m(J^+_e,J^-_e)=J^+_e+O(\epsilon_{\mathrm{eq}})
\end{align}
because any homogeneous symmetric mean $m$ is bounded as $\min(a,b)\leq m(a,b)\leq\max(a,b)$. Thus, the TSL~\eqref{TSLs} reduces to
\begin{align}
    \langle\sigma\rangle_{\tau}\geq\frac{\langle v_1\rangle_{\tau}^2}{\langle\sum_{e\in\mathcal{E}}|J_e^+|\rangle_{\tau}},
    \label{TSL eq}
\end{align}
which is independent of $m$. Here, the hierarchy of the TSLs arises from the higher-order terms of $\epsilon_{\mathrm{eq}}$.

In Fig.~\ref{fig:TL}, we numerically demonstrate the above facts using the MJP with three microstates $\mathcal{S}=\{1,2,3\}$ and one heat reservoir $\mathcal{R}=\{1\}$. We set the transition rates as $R^{(1)}_{21}=R^{(1)}_{32}=R^{(1)}_{13}=\mathrm{e}^{F_C/6}$ and $R^{(1)}_{12}=R^{(1)}_{23}=R^{(1)}_{31}=\mathrm{e}^{-F_C/6}$. Here, $F_C$ corresponds to the cycle affinity~\cite{schnakenberg1976network,seifert2012stochastic}. The steady state is given by $x^{\mathrm{st}}\coloneqq(1/3,1/3,1/3)^{\top}$ regardless of $F_C$. This steady state becomes an equilibrium state only when $F_C$ is zero. In the following calculations, we consider the following four cases: (I) far from equilibrium ($F_C=0$, $x(0)=(0.01,0.1,0.89)^{\top}$), (II) near equilibrium ($F_C=0$, $x(0)=(0.3,0.38,0.32)^{\top}$), (III) far from the nonequilibrium steady state ($F_C=3$, $x(0)=(0.01,0.1,0.89)^{\top}$), and (IV) near the nonequilibrium steady state ($F_C=3$, $x(0)=(0.3,0.38,0.32)^{\top}$).

First, we focus on the cases of $F_C=0$, where the system relaxes to the equilibrium. In case (I), the distribution evolves fast as shown in Fig.~\ref{fig:TL}(a). Since the speed of the time evolution is not small, a smaller mean does not necessarily provide a tighter bound as shown in Fig.~\ref{fig:TL}(b) and (c). For example, the contraharmonic mean yields the weakest bound even though this mean is smaller than the maximum. The harmonic, geometric, and logarithmic means provide tighter bounds than the one based on the smallest mean, the minimum. Although the ordering of the bounds can vary depending on $t$ as shown in Fig.~\ref{fig:TL}(c), we can confirm the hierarchy between the TSLs based on $\min$ and $\max$ in Eq.~\eqref{hierarchy minmax}. In contrast to case (I), the speed of the time evolution is slow in case (II) [Fig.~\ref{fig:TL}(d)]. In case (II), the bounds take almost the same values as shown in Fig.~\ref{fig:TL}(e), since the system is near equilibrium. We can see the hierarchy in Fig.~\ref{fig:TL}(f): a smaller mean provides a tighter bound, since $v_1$ is slow.

Second, we focus on the cases of $F_C=3$, where the system approaches the nonequilibrium steady state. Figure~\ref{fig:TL}(g-i) shows the details of case (III). This case is similar to case (I) because the speed of the time evolution is not small. A smaller mean does not necessarily provide a tighter bound, and the ordering of the bounds can vary depending on $t$. The hierarchy between the TSLs based on $\min$ and $\max$ in Eq.~\eqref{hierarchy minmax} also holds as shown in Fig.~\ref{fig:TL}(i). In case (IV), the speed of the time evolution is slow as in case (II) [Fig.~\ref{fig:TL}(j)]. In this case, the bounds take different values as shown in Fig.~\ref{fig:TL}(k). This is because the system is not near equilibrium. We can also see the hierarchy between TSLs in Fig.~\ref{fig:TL}(l), since $v_1$ is slow.

\subsection{Minimum dissipation and achievability of equality in TSLs}
\label{sec:minimum dissipation}

As in the special cases~\cite{dechant2022minimum,van2023thermodynamic}, the lower bound of the dissipation in Eq.~\eqref{TSLs2} provides a minimum dissipation formula. Under the optimal protocol that achieves the minimum dissipation, \add{both} lower bounds in Eq.~\eqref{TSLs2} coincide with $\langle\sigma\rangle_{\tau}$; in other words, the equalities in the TSLs are achieved.


We show these facts by considering a minimum dissipation required to evolve the state from $x(0)$ to $x(\tau)$ over time $\tau$ under some physically valid conditions. In the following, we regard the EP and the time average of the activity as functionals of the fluxes as
\begin{empheq}[left=\empheqlbrace]{align}
    \Sigma_{\tau}[J^+, J^-]&\coloneqq\int_{0}^{\tau}dt\sum_{e\in\mathcal{E}}(J^+_e-J^-_e)\ln\frac{J^+_e}{J^-_e},\\
    \langle\mu_{m}\rangle_{\tau}[J^+, J^-]&\coloneqq\frac{1}{\tau}\int_{0}^{\tau}dt\sum_{e\in\mathcal{E}}m(J^+_e,J^-_e).
\end{empheq}

We consider the minimization problem $\inf_{J^+, J^-}\Sigma_{\tau}[J^+, J^-]$ under the following conditions: (i) the fluxes evolve the distribution from $x(0)$ to $x(\tau)$ as
\begin{align}
    x(\tau)-x(0)=\nabla^{\top}\int_{0}^{\tau}dt\left(J^+-J^-\right),
\end{align}
(ii) the time average of the activity is bounded by a constant $M_0$ as
\begin{align}
    \langle\mu_{m}\rangle_{\tau}[J^+, J^-]\leq M_0,
    \label{condition_M0}
\end{align}
(iii) if $J^+_e=J^-_e=0$ holds, we regard $(J^+_e-J^-_e)\ln(J^+_e/J^-_e)$ as zero. Condition (iii) is physically valid since $J^+_e=J^-_e=0$ implies that jump does not occur on edge $e$ for MJPs and reaction $e$ does not proceed for CRNs. Then, the minimum dissipation is related to the 1-Wasserstein distance as 
\begin{align}
    &\inf_{J^+, J^-}\Sigma_{\tau}[J^+, J^-]\nonumber\\
    &=W_1(x(0),x(\tau))\Psi_m^{-1}\left(\frac{W_1(x(0),x(\tau))}{\tau M_0}\right).
    \label{minimum_EP}
\end{align}
We remark that we can make the EP zero by allowing the activity to have an infinite value~\cite{dechant2022minimum}. We impose condition (ii) to prevent such non-physical optimization. We provide the derivation of the minimum dissipation formula~\eqref{minimum_EP} in Appendix~\ref{ap:minimum dissipation}.

As an optimizer of this minimization problem, we can take the fluxes $J^{+\star}$ and $J^{-\star}$ that satisfy the following properties: (a) The current generated by $J^{+\star}$ and $J^{-\star}$, i.e., $J^{\star}\coloneqq J^{+\star}-J^{-\star}$, is independent of time. (b) There exists a potential $\tilde{\varphi}$ that satisfies
\begin{align}
    \ln\frac{J^{+\star}_e}{J^{-\star}_e}=-(\nabla\tilde{\varphi})_e,
\end{align}
for all $e$ where $J^{\pm\star}_e\neq0$. Property (a) determines the concrete form of the distribution evolved by the optimal current $J^{\star}$ from $x(0)$, denoted by $x^{\star}$, as 
\begin{align}
    x^{\star}(t)\coloneqq\left(1-\frac{t}{\tau}\right)x(0)+\frac{t}{\tau}x(\tau).
\end{align}
It is easily verified by solving $d_tx^{\star}=\nabla^{\top}J^{\star}$ with the initial condition $x^{\star}(0)=x(0)$.
Property (b) implies that the force generated by $J^{\pm\star}$ can be regarded as a conservative force after removing the edges that satisfy $J^{\pm\star}_e=0$. We note that this removal of edges does not affect the time evolution and dissipation. This conservativeness is based on the Kantorovich--Rubinstein duality. We provide further details of the optimizers $J^{+\star}$ and $J^{-\star}$ in Appendix~\ref{ap:minimum dissipation}.

For special cases ($m=A,L$), previous studies show that the optimal force can be realized by a conservative force~\cite{dechant2022minimum,van2023thermodynamic}. Generally, whether a conservative force can achieve the minimum EP in MJPs depends on the conditions imposed on the minimization problem~\cite{remlein2021optimality,ilker2022shortcuts, yoshimura2023housekeeping}. Our results show that under the constraint in Eq.~\eqref{condition_M0}, the minimum EP can always be achieved by a conservative force, regardless of which activity is used.

The minimum dissipation~\eqref{minimum_EP} ensures that the lower bound in Eq.~\eqref{TSLs2} is achievable. Then the two lower bounds in Eqs.~\eqref{TSLs} and~\eqref{TSLs2} coincide. Thus, the TSLs can be seen as achievable bounds. It is remarkable that in the optimal situation, the EPR also becomes time independent since the optimizer is, as discussed.


\section{Comparison with excess entropy production rates and TSL for 2-Wasserstein distance}
\label{sec:excess comparison}

\add{In the above discussion, the same approach applies to all choices of means. The TSLs can be obtained regardless of which means are used to define the activities as long as they satisfy the conditions in Eqs.~\eqref{condition_fM'} and~\eqref{condition_fM''}. In this section, in contrast to the previous sections, we examine cases where some particular means play a critical role.}

The TSLs are inequalities about the speed of the time evolution. This fact implies that TSLs are related to the nonstationarity of the system. Therefore, some TSLs can be tightened by replacing the EPR with the excess EPR~\cite{oono1998steady}, which may be the nonstationary contribution of the EPR.

The definition of the excess EPR is not unique~\cite{hatano2001steady,maes2014nonequilibrium}. A typical example is the Hatano--Sasa excess (or nonadiabatic) EPR~\cite{hatano2001steady,esposito2010three}. This method is based on the steady state, and applicable to limited systems: Markov processes and a special class of CRNs, which is called complex balanced CRNs~\cite{rao2016nonequilibrium}. To consider more general systems, the geometric excess/housekeeping decomposition for Langevin systems~\cite{dechant2022geometric_E,dechant2022geometric_R,ito2023geometric} has been extended based on different geometries: optimal transport theory~\cite{yoshimura2023housekeeping}, information geometry~\cite{kolchinsky2022information}, and Hessian geometry~\cite{kobayashi2022hessian}.
Especially, the excess EPR based on optimal transport theory~\cite{yoshimura2023housekeeping} is related to the 2-Wasserstein distance~\cite{maas2011gradient}, which is introduced by the gradient structure of dynamics.

In this section, we compare the lower bound of $\langle\sigma\rangle_{\tau}$ in Eq.~\eqref{TSLs} with excess EPRs. We focus on the geometric excess EPRs in Refs.~\cite{yoshimura2023housekeeping,kolchinsky2022information}, since several TSLs that constrain them using the 1-Wasserstein distance have already been discovered. We also compare the TSL in Eq.~\eqref{TSLs} with the TSL for another distance based on optimal transport theory, i.e., the 2-Wasserstein distance~\cite{maas2011gradient,yoshimura2023housekeeping}. \add{We clarify the following point through analytical and numerical comparison. The lower bounds on the total EPR by the TSLs based on the arithmetic or logarithmic mean also bound the excess EPRs. However, some TSLs with other means do not necessarily provide lower bounds of excess EPRs. This implies that certain means should be used to bound the excess EPRs, in contrast to the case of total EPR.}

\subsection{Geometric excess EPRs}
We introduce two types of geometric excess EPRs, which are nonnegative lower bounds on the EPR. The first is the \textit{Onsager geometric} excess EPR $\sigma^{\mathrm{ONS}}_{\mathrm{ex}}$~\cite{yoshimura2023housekeeping}, which satisfies $0 \leq \sigma^{\mathrm{ONS}}_{\mathrm{ex}} \leq \sigma$. Based on the form of the EPR in Eq.~\eqref{rewrite EPR} with $m=L$, this quantity is defined as
\begin{align}
\sigma^{\mathrm{ONS}}_{\mathrm{ex}}\coloneqq\min_{\hat{J}}\sum_{e\in\mathcal{E}}\frac{\hat{J}_e^{2}}{\mu_{L,e}}.
    \label{onsager excess}
\end{align}
Here, the minimization is performed over all currents $\hat{J}$ that reproduces the original time evolution at the moment as
\begin{align}
    d_t x(t)=\nabla^{\top}\hat{J}.
    \label{condition:onsager excess}
\end{align}
\add{This condition implies that $\sigma^{\mathrm{ONS}}_{\mathrm{ex}}$ is the minimum dissipation rate required to achieve the original time evolution. We note that $\sigma^{\mathrm{ONS}}_{\mathrm{ex}}$ is zero when the system is in steady state. This is verified as follows: Let $J^{\mathrm{ONS}}$ denote the minimizer of the right-hand side of Eq.~\eqref{onsager excess}. In steady state, the condition~\eqref{condition:onsager excess} becomes $0=\nabla^{\top}\hat{J}$. Thus, we can make $\sigma^{\mathrm{ONS}}_{\mathrm{ex}}$ vanish by taking $J^{\mathrm{ONS}}=0$, which certainly satisfies $0=\nabla^{\top}J^{\mathrm{ONS}}$.}

Reference~\cite{kolchinsky2022information} provides another definition of the excess EPR, information geometric excess EPR.
This is based on the form of the EPR with the KL divergence~\eqref{KL form EPR} and defined as
\begin{align}
    \sigma^{\mathrm{IG}}_{\mathrm{ex}}\coloneqq\min_{\hat{J}^{+}, \hat{J}^{-}{}}\left\{D_{\mathrm{KL}}(\hat{J}^{+}\|J^{-})+D_{\mathrm{KL}}(\hat{J}^{-}\|J^{+})\right\}.
    \label{information geometric excess}
\end{align}
Here, the minimization is performed over all fluxes $\hat{J}^{+}$ and $\hat{J}^{-}$ that reproduces the original time evolution at the moment as
\begin{align}
    d_t x(t)=\nabla^{\top}(\hat{J}^{+}-\hat{J}^{-}).
    \label{condition: info}
\end{align}
This information geometric excess EPR also satisfies $0 \leq \sigma^{\mathrm{IG}}_{\mathrm{ex}} \leq \sigma$. We remark that there is a hierarchy between these two excess EPRs as~\cite{kolchinsky2022information}
\begin{align}
    \sigma^{\mathrm{ONS}}_{\mathrm{ex}}\geq\sigma^{\mathrm{IG}}_{\mathrm{ex}}.
    \label{excess hierarchy}
\end{align}
\add{Due to the condition in Eq.~\eqref{condition: info}, we may regard $\sigma^{\mathrm{IG}}_{\mathrm{ex}}$ as the minimum dissipation rate required to achieve the original time evolution. The information geometric excess EPR also vanishes when the system is in steady state, as shown below. Let $J^{+\mathrm{IG}}$ and $J^{-\mathrm{IG}}$ denote the minimizer of the right-hand side of Eq.~\eqref{information geometric excess}. In steady state, the condition in Eq.~\eqref{condition: info} becomes $0=\nabla^{\top}(\hat{J}^{+}-\hat{J}^{-})$. We obtain $\sigma^{\mathrm{IG}}_{\mathrm{ex}}=0$ by choosing $J^{+\mathrm{IG}}=J^{-}$ and $J^{-\mathrm{IG}}=J^{+}$, since they satisfy the condition in Eq.~\eqref{condition: info} as $\nabla^{\top}(J^{+\mathrm{IG}}-J^{-\mathrm{IG}})=-\nabla^{\top}(J^{+}-J^{-})=0$. Here, we use the fact that $\nabla^{\top}(J^{+}-J^{-})=d_tx=0$ holds in steady state.}

\add{
We note that we can also obtain these excess EPRs by minimization problems with forces, whose optimal solutions are conservative forces~\cite{yoshimura2023housekeeping,kolchinsky2022information}. Thus, these excess EPRs correspond to the minimum dissipation rate achieved by the conservative forces under different conditions. In other words, these excess EPRs measure the conservativeness and nonstationarity of the system.
}

\subsection{2-Wasserstein distance}

In Ref.~\cite{yoshimura2023housekeeping}, the authors use another Wasserstein distance, that is, the 2-Wasserstein distance, to derive a TSL for the Onsager excess EPR. 

Here, we introduce the 2-Wasserstein distance. Since the fluxes depend on $x$, the activity also depends on $x$. We write $\mu_{m,e}(x)$ to indicate this dependence on $x$. If $\mu_{L,e}(x)$ is independent of $t$, we can define the 2-Wasserstein distance as
\begin{align}
    W_2(x^A,x^B)\coloneqq\sqrt{\inf_{\hat{x},\hat{J}}\tau\int_{0}^{\tau}dt\sum_{e\in\mathcal{E}}\frac{\hat{J}_e^{2}}{\mu_{L,e}(\hat{x})}},
    \label{2-Wasserstein distance}
\end{align}
where $\hat{x}$ and $\hat{J}$ satisfy
\begin{align}
    \hat{x}(0)=x^A,\;\hat{x}(\tau)=x^B,\;d_t\hat{x}(t)=\nabla^{\top}\hat{J}(t).
    \label{conditions for 2-Wasserstein distance}
\end{align}
This definition generalizes a formulation of the 2-Wasserstein distance for probability distributions with continuous variables, so called the Benamou--Brenier formula~\cite{benamou2000computational}. Equation~\eqref{2-Wasserstein distance} is introduced for detailed-balanced MJPs in Refs.~\cite{maas2011gradient,chow2012fokker,mielke2013geodesic}, and extended to more general systems in Ref.~\cite{yoshimura2023housekeeping}.


We can measure the speed of the time evolution with the 2-Wasserstein distance as
\begin{align}
    v_2(t)\coloneqq\lim_{\varDelta t\to0}\frac{W_2(x(t),x(t+\varDelta t))}{\varDelta t}.
\end{align}
In contrast to the 2-Wasserstein distance itself, we can define this speed even though $\mu_{L,e}(x)$ depends on time, since we can regard $\mu_{L,e}(x)$ as a constant in the infinitesimal time interval. We can relate the speed $v_2$ to the Onsager geometric excess EPR as
\begin{align}
    \sigma^{\mathrm{ONS}}_{\mathrm{ex}}=v_2^2.
    \label{onsager excess speed}
\end{align}
This is easily verified as below. Taking $\varDelta t\ll 1$, we can reduce the definition of the 2-Wasserstein distance~\eqref{2-Wasserstein distance} as
\begin{align}
    W_2(x(t),x(t+\varDelta t))^2\coloneqq\inf_{\hat{J}}\varDelta t^2\sum_{e\in\mathcal{E}}\frac{\hat{J}_e^{2}}{\mu_{L,e}(x)}+O(\varDelta t^3).
    \label{reduced 2-Wasserstein distance}
\end{align}
Here, the conditions~\eqref{conditions for 2-Wasserstein distance} make $\hat{J}$ satisfy
\begin{align}
    x(t+\varDelta t)-x(t)=\varDelta t(\nabla^{\top}\hat{J})+O(\varDelta t^2).
    \label{reduced conditions for 2-Wasserstein distance}
\end{align}
Since this condition is equivalent to Eq.~\eqref{condition:onsager excess} in the limit $\varDelta t\to0$, we obtain Eq.~\eqref{onsager excess speed} by dividing the both sides of Eq.~\eqref{reduced 2-Wasserstein distance} and taking the limit $\varDelta t\to0$.

\subsection{TSLs for geometric excess EPRs}

Here, we introduce some TSLs for the geometric excess EPRs and the Wasserstein distances. 

The relation in Eq.~\eqref{onsager excess speed} yields a TSL for the Onsager geometric excess EPR and the 2-Wasserstein distance~\cite{yoshimura2023housekeeping} as
\begin{align}
    \langle\sigma^{\mathrm{ONS}}_{\mathrm{ex}}\rangle_{\tau}\geq\langle v_2\rangle_{\tau}^2.
    \label{TSL2 ONS}
\end{align}
This is easily verified by taking the time integration of Eq.~\eqref{onsager excess speed} and using the Cauchy--Schwarz inequality as $\int_{0}^{\tau}dt\,\sigma^{\mathrm{ONS}}_{\mathrm{ex}}=\int_{0}^{\tau}dt\, v_2^2=(\int_{0}^{\tau}dt\,1^2\int_{0}^{\tau}dt\, v_2^2)/\tau\geq(\int_{0}^{\tau}dt\,v_2)^2/\tau$. Since the Onsager geometric excess EPR is smaller than or equivalent to the EPR, we also obtain $\langle\sigma\rangle_{\tau}\geq\langle v_2\rangle_{\tau}^2$. This is a generalization of TSLs for Langevin systems studied in Refs.~\cite{aurell2011optimal,aurell2012refined,nakazato2021geometrical}

We can also obtain the TSL for the Onsager geometric excess EPR and the 1-Wasserstein distance as
\begin{align}
    \langle\sigma^{\mathrm{ONS}}_{\mathrm{ex}}\rangle_{\tau}\geq\frac{\langle v_1\rangle_{\tau}^2}{\langle\mu_L\rangle_{\tau}}.
    \label{TSL1L ONS}
\end{align}
This is derived for a more general set-up in Ref.~\cite{nagayama2023geometric}. We also provide the proof in Appendix~\ref{ap:derivation ONS TSL}. We can obtain the TSL [Eq.~\eqref{TSLs}] for the logarithmic mean $m=L$ by combining Eq.~\eqref{TSL1L ONS} and the inequality $\langle \sigma \rangle_{\tau} \geq \langle\sigma^{\mathrm{ONS}}_{\mathrm{ex}} \rangle_{\tau}$. 

Reference~\cite{kolchinsky2022information} also provides the TSL for the information geometric excess EPR and the 1-Wasserstein distance as
\begin{align}
    \langle\sigma^{\mathrm{IG}}_{\mathrm{ex}}\rangle_{\tau}\geq2\langle v_1\rangle_{\tau}\tanh^{-1}\left(\frac{\langle v_1\rangle_{\tau}}{2\langle\mu_A\rangle_{\tau}}\right),
    \label{TSL1A IG}
\end{align}
and thus we can obtain the TSL [Eq.~\eqref{TSLs}] for the arithmetic mean $m=A$ from the inequality $\langle \sigma \rangle_{\tau} \geq \langle\sigma^{\mathrm{IG}}_{\mathrm{ex}} \rangle_{\tau}$.
Combining this TSL and the hierarchy in Eq.~\eqref{excess hierarchy}, we also obtain another TSL for the Onsager geometric excess EPR and the 1-Wasserstein distance as
\begin{align}
    \langle\sigma^{\mathrm{ONS}}_{\mathrm{ex}}\rangle_{\tau}\geq2\langle v_1\rangle_{\tau}\tanh^{-1}\left(\frac{\langle v_1\rangle_{\tau}}{2\langle\mu_A\rangle_{\tau}}\right).
    \label{TSL1A ONS}
\end{align}

The above three TSLs can be recast in terms of $\Psi_m$. The TSLs in Eqs.~\eqref{TSL1L ONS} and ~\eqref{TSL1A ONS} imply that $\langle v_1\rangle_{\tau}\Psi_m^{-1}(\langle v_1\rangle_{\tau}/\langle \mu_m\rangle_{\tau})$ becomes a lower bound of the time average of the Onsager geometric excess EPR in the cases of $m=L$ and $m=A$. The TSL in Eq.~\eqref{TSL1A IG} implies that $\langle v_1\rangle_{\tau}\Psi_m^{-1}(\langle v_1\rangle_{\tau}/\langle \mu_m\rangle_{\tau})$ becomes a lower bound of the time average of the information geometric excess EPR in the case of $m=A$. 
Consequently, the TSLs in Eqs.~\eqref{TSL1L ONS}, \eqref{TSL1A IG}, and ~\eqref{TSL1A ONS} tighten the lower bound for the total EPR given in Eq.~\eqref{TSLs} by the excess EPRs. 

\begin{figure}
    \centering
    \includegraphics[width=\linewidth]{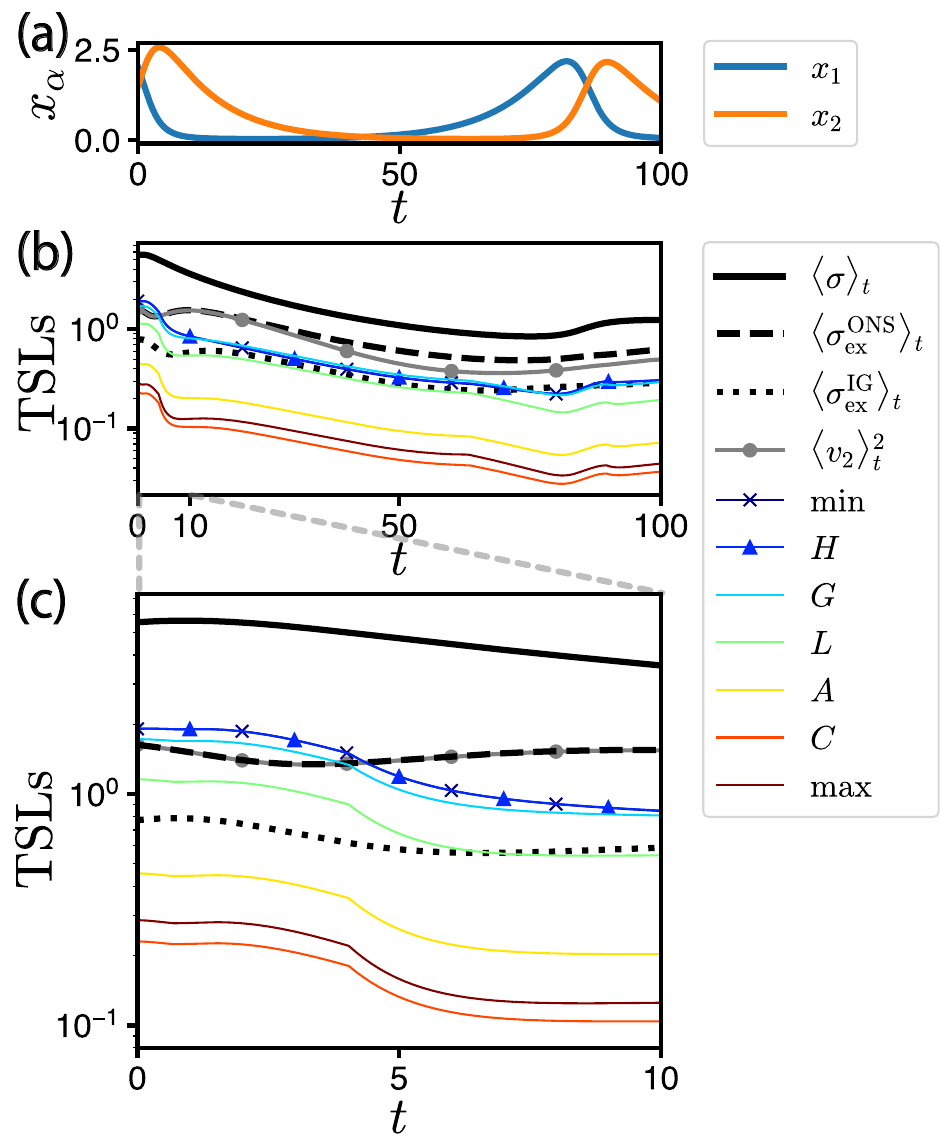}
    \caption{Comparison of the TSLs [Eq.~\eqref{TSLs}], the TSL for the 2-Wasserstein distance [Eq.~\eqref{TSL2 ONS}], and the geometric excess EPRs. We use the damped Lotka--Volterra chemical reaction model. (a) The time series of concentration distribution. (b) The TSLs and the geometric excess EPRs. The symbols in the legend indicate the means used to obtain the TSLs, which are the same as those listed in TABLE.~\ref{tab:means}. (c) An enlarged view of (b) for the time period $[0, 10]$. Note that the TSLs based on $m=\min$ and $m=H$ overlap as indicated by the markers. The lower bounds $\langle v_1\rangle_{\tau}\Psi_m^{-1}(\langle v_1\rangle_{\tau}/\langle\mu_m\rangle_{\tau})$ for $m=\min, H, G$ can be greater than $\langle\sigma^{\mathrm{IG}}_{\mathrm{ex}}\rangle_{\tau}$, $\langle\sigma^{\mathrm{ONS}}_{\mathrm{ex}}\rangle_{\tau}$, and $\langle v_2\rangle_{\tau}$. In addition, the lower bound for $m=L$ can be greater than $\langle\sigma^{\mathrm{IG}}_{\mathrm{ex}}\rangle_{\tau}$.}
    \label{fig:LV}
\end{figure}

\subsection{Numerical comparison}


Let us consider more deeply the connection between the excess EPRs and general bounds. 
First, we can ask what if the mean in $\langle v_1\rangle_{\tau}\Psi_m^{-1}(\langle v_1\rangle_{\tau}/\langle \mu_m\rangle_{\tau})$ is not $A$ or $L$. 
In addition, it is not certain which is tighter, the TSL~\eqref{TSL2 ONS}, which involves the 2-Wasserstein-based speed $v_2$, and the TSL~\eqref{TSLs}, containing the 1-Wasserstein speed measure $v_1$. 
In this section, we examine these questions numerically. 
In particular, we compare the geometric excess EPRs, the TSL for the 2-Wasserstein distance~\eqref{TSL2 ONS}, and our TSLs in TABLE.~\ref{tab:TSLs}.

Here, we use the following damped Lotka--Volterra chemical reaction model~\cite{strogatz2018nonlinear}:
\begin{align}
    \ce{$X_1 + A$ <=>C[\kappa^+_1][\kappa^-_1] $2X_1$},\;\ce{$X_1 + X_2$ <=>C[\kappa^+_2][\kappa^-_2] $2X_2$},\;\ce{$X_2$ <=>C[\kappa^+_3][\kappa^-_3] $B$}.
    \label{LV model}
\end{align}
Here, we label the reactions as $e=1$, $2$, and $3$ from left to right in Eq.~\eqref{LV model}. We also assume the mass action kinetics and $\kappa^{\pm}_e$ is the reaction rate constant for reaction $e$. The species $A$ and $B$ are the external ones and their concentrations are fixed as $1$. We set the reaction rate constants as $(\kappa^+_1,\kappa^+_2,\kappa^+_3,\kappa^-_1,\kappa^-_2,\kappa^-_3)=(0.1,0.2,0.1,10^{-3},10^{-3},10^{-5})$. Then, the time evolution of the concentration $x=(x_1,x_2)^{\top}$ is given by the following rate equation:
\begin{align}
\left\{
\begin{aligned}
&d_t x_1=0.1x_1-10^{-3}x_1^2-0.2x_1x_2+10^{-3}x_2^2\\
&d_t x_2=0.2x_1x_2-10^{-3}x_2^2-0.1x_2+10^{-5} 
\end{aligned}
\right..\label{LV rate eq}
\end{align}
To demonstrate the TSLs, we use the time series of $x$ obtained by solving the rate equation~\eqref{LV rate eq} with the initial condition $x(0)=(2,1.5)^{\top}$. This time series is shown in Fig.~\ref{fig:LV}(a).

In Fig.~\ref{fig:LV}(b), we show the comparison of the time-averaged geometric excess EPRs and the lower bounds provided by the TSLs. We can confirm that the TSLs for the geometric excess EPRs in Eqs.~\eqref{TSL2 ONS}, ~\eqref{TSL1L ONS}, ~\eqref{TSL1A ONS}, and ~\eqref{TSL1A IG} hold. However, the inequality may no longer hold when the combination of the excess EPR and the lower bound (or the mean used) is changed. For example, $\langle v_2\rangle_{\tau}^2$ becomes larger than $\langle\sigma^{\mathrm{IG}}_{\mathrm{ex}}\rangle_{\tau}$ in contrast to Eq.~\eqref{TSL2 ONS}. We can also confirm that $\langle v_1\rangle_{\tau}\Psi_m^{-1}(\langle v_1\rangle_{\tau}/\langle\mu_m\rangle_{\tau})$ can exceed the time-averaged excess EPRs for some $m$. Let us focus on $t\in[0,10]$ shown in Fig.~\ref{fig:LV}(c). In this period of time, the lower bound $\langle v_1\rangle_{\tau}\Psi_m^{-1}(\langle v_1\rangle_{\tau}/\langle\mu_m\rangle_{\tau})$ for $m=\min,H,G$ can be greater than $\langle\sigma^{\mathrm{IG}}_{\mathrm{ex}}\rangle_{\tau}$ and $\langle\sigma^{\mathrm{ONS}}_{\mathrm{ex}}\rangle_{\tau}$. The lower bound $\langle v_1\rangle_{\tau}\Psi_L^{-1}(\langle v_1\rangle_{\tau}/\langle\mu_L\rangle_{\tau})$ also becomes larger than $\langle\sigma^{\mathrm{IG}}_{\mathrm{ex}}\rangle_{\tau}$. These observations imply that we need to use specific means to bounce geometric excess EPRs. \add{This contrasts with total EPR, for which any mean can be used to obtain a lower bound.}

It is remarkable that some TSLs based on the 1-Wasserstein distance can be tighter than the one based on the 2-Wasserstein distance. In Fig.~\ref{fig:LV}(b), $\langle v_2\rangle_{\tau}^2$ becomes larger than the lower bounds based on $\langle v_1\rangle_{\tau}$ at most times. However, in Fig.~\ref{fig:LV}(c), the TSLs~\eqref{TSLs} for $m=\min, H, G$ are possibly tighter than the TSL for the 2-Wasserstein distance. This is different from the case of Langevin systems, where the TSL based on the 1-Wasserstein distance is always weaker than the one based on the 2-Wasserstein distance~\cite{nagayama2023geometric}.


\section{Discussion}

In this paper, we have derived an infinite variety of TSLs based on the 1-Wasserstein distance. The TSLs can be applied to MJPs and CRNs. We have also related the lower bound of EP provided by each TSL to the minimum dissipation. This minimum dissipation is achievable with a conservative force.

Let us discuss the choice of the mean. As confirmed through numerical calculations, there is no apparent hierarchy among the TSLs. Thus, the mean providing the tightest TSL depends on the situation. Moreover, the equalities in the TSLs are achievable by minimizing dissipation under the constraint of the time-averaged activity measured with the respective mean. This implies that the previously known TSLs based on the 1-Wasserstein distance~\cite{dechant2022minimum,van2023thermodynamic, nagase2024thermodynamically} are not necessarily the best bounds for the time-averaged EPR.

The choice of the mean becomes crucial for obtaining a lower bound of the time-averaged excess EPRs. This may be because the excess EPRs are defined by optimization problems based on specific means~\cite{yoshimura2023housekeeping,kobayashi2022hessian}. Therefore, like the TSLs in this work, we may understand the arbitrariness in the definition of excess EPRs~\cite{yoshimura2023housekeeping,kolchinsky2022information,kobayashi2022hessian,kobayashi2023information} from the perspective of means.

Furthermore, it may be possible to characterize an intrinsic lower bound of the time-averaged EPR based on our results. In fact, by selecting the mean that provides the tightest bound at each time, one could obtain a tighter bound than the existing TSLs. The mean that provides the tightest bound may also reflect the characteristics of the dynamics. Hence, it is an interesting challenge to explore the correspondence between the nature of the dynamics and the mean that gives the tightest TSL.

We also discuss the relationship between our results and thermodynamic uncertainty relations (TURs), which describe trade-off relations between the EP and the accuracy~\cite{barato2015thermodynamic,horowitz2020thermodynamic}. As in the case of TSLs, TURs are also derived in various set-ups using different ways~\cite{Gingrich2016,horowitz2017proof,maes2017frenetic,proesmans2017discrete,PietzonkaFT2017,Dechantccurrent2018,DechantMulti2018,TimpanaroEFT2019,potts2019thermodynamic,Hasegawa2019,DechantImproving2021,DechantFRI2020,otsubo2020estimating,Liu2020,dechant2022geometric_E,yoshimura2023housekeeping,kwon2024unified}, and some of them correspond to the means: The arithmetic mean is used to unify the TURs and the TSLs based on the dynamical activity~\cite{Falasco2020,vo2020unified,Vo2022,Hasegawa2023}. The geometric mean also appears in the initial derivation of TURs~\cite{Gingrich2016,horowitz2017proof}. This is because the geometric mean is essential to obtain the rate function, which quantifies fluctuations for the empirical current and distribution~\cite{Maes2008,Barato2015}. The logarithmic mean is used to derive the TURs for the Onsager geometric excess and housekeeping EPRs~\cite{yoshimura2023housekeeping}. This implies that using a different mean may lead to a different TUR as in the case of the TSLs.

Finally, we introduce some future directions: one direction is to extend our results to open quantum systems. Historically, TSLs for MJPs have been extended to open quantum systems described by the quantum master equation~\cite{Funo2019,van2021geometrical}. In particular, reference~\cite{van2023thermodynamic} considers the quantum extension of the second line of the TSLs~\eqref{TSLs} for $m=A, L$. Reference~\cite{yoshimura2024force} also develops a quantum version of the relation between force and current in Eq.~\eqref{relation FtoJ} in the case of $m=L$. Extending the TSLs and the force-current relation to quantum systems with other $m$ remains an open challenge.

The other direction is to elucidate the relationship between our TSLs and classical information-geometric speed limits (ISLs) based on the Fisher information~\cite{salamon1983thermodynamic,crooks2007measuring, sivak2012thermodynamic, ito2018stochastic} and the Cram\'{e}r-Rao bound~\cite{ito2020stochastic,nicholson2020time,yoshimura2021information}. There is a mathematical similarity between the classical ISLs and the TSLs for the 2-Wasserstein distance~\cite{nakazato2021geometrical,li2023wasserstein,ito2023geometric,nagayama2023geometric,chennakesavalu2023unified, zhong2024beyond}. However, clarifying the similarities and differences between ISLs and the TSLs for the 1-Wasserstein distance is still an open question. Moreover, the classical ISLs originate from the quantum speed limit for quantum systems~\cite{mandelstam1991uncertainty,anandan1990geometry} using the quantum Fisher information~\cite{ pires2016generalized,GarcaPintos2022}. Interestingly, we can define the quantum Fisher information using various means as in the case of the general activity~\cite{petz2011introduction}. Investigating the connection between our TSLs for classical systems and ISLs for quantum systems~\cite{bettmann2024information} would also be an intriguing direction.

\begin{acknowledgments}
    Authors thank Artemy Kolchinsky and Naruo Ohga for their suggestive comments. R.N. also \ thanks Yasushi Okada for their suggestive comments. 
    K.Y.\ is supported by Grant-in-Aid for JSPS Fellows (Grant No.~22J21619). 
    S.I.\ is supported by JSPS KAKENHI Grants No.~21H01560, No.~22H01141, No.~23H00467, and No.~24H00834, 
    JST ERATO Grant No.~JPMJER2302, 
    and UTEC-UTokyo FSI Research Grant Program.
    This research was supported by JSR Fellowship, the University of Tokyo. 
\end{acknowledgments}

\appendix

\section{The equivalence of the detailed balance condition and the conservativeness}
\label{ap:conservativeness}

Here we prove the equivalence of the following two statements in the CRN with mass action kinetics and the MJP: (i) the system satisfies the detailed balance condition at time $t$, and (ii) the forces are conservative at time $t$. 

We only consider the CRN with mass action kinetics, since we can regard the MJP as a special case of the CRN with mass action kinetics. This identification is done by replacing $(n_{\alpha e}^{+},n_{\alpha e}^{-}, \kappa_e^{+}, \kappa_{e}^{-})$ in the CRN with $\left(\delta_{\alpha \mathrm{s}(e)},\delta_{\alpha \mathrm{t}(e)}, R_{\mathrm{t}(e)\mathrm{s}(e)}^{(\mathrm{r}(e))}, R_{\mathrm{s}(e)\mathrm{t}(e)}^{(\mathrm{r}(e))}\right)$. 

We also remark that the mass action kinetics rewrites the definition of force~\eqref{thermodynamic force} as
\begin{align}
    F_e(t)&=\ln\frac{\kappa_{e}^{+}(t)\prod_{\alpha\in\mathcal{S}}[x_{\alpha}(t)]^{n_{\alpha e}^{+}}}{\kappa_{e}^{-}(t)\prod_{\alpha\in\mathcal{S}}[x_{\alpha}(t)]^{n_{\alpha e}^{-}}}\notag\\
    &=\ln\frac{\kappa_{e}^{+}(t)}{\kappa_{e}^{-}(t)}-\sum_{\alpha\in\mathcal{S}}\nabla_{e\alpha}\ln x_{\alpha}(t).
    \label{Force_mass action}
\end{align}

First, we prove (i)$\Rightarrow$(ii). Due to the detailed balance condition, their exist the equilibrium state $x^{\mathrm{eq}}(t)$ that satisfies
\begin{align}
    0=J_e(x^{\mathrm{eq}}(t),(t))=J_e^+(x^{\mathrm{eq}}(t),(t))-J_e^-(x^{\mathrm{eq}}(t),(t)),
\end{align}
for all $e\in\mathcal{E}$. Thus, the equilibrium state satisfies $\ln [J_e^+(x^{\mathrm{eq}}(t),(t))/J_e^-(x^{\mathrm{eq}}(t),(t))]=0$. The mass action kinetics rewrites this as
\begin{align}
    \ln\frac{\kappa_{e}^{+}(t)}{\kappa_{e}^{-}(t)}&=\ln\frac{\prod_{\alpha\in\mathcal{S}}[x^{\mathrm{eq}}_{\alpha}(t)]^{n_{\alpha e}^{-}}}{\prod_{\alpha\in\mathcal{S}}[x^{\mathrm{eq}}_{\alpha}(t)]^{n_{\alpha e}^{+}}}\notag\\
    &=\sum_{\alpha\in\mathcal{S}}(n_{\alpha e}^{-}-n_{\alpha e}^{+})\ln x^{\mathrm{eq}}_{\alpha}(t)\notag\\
    &=\sum_{\alpha\in\mathcal{S}}\nabla_{e\alpha}\ln x^{\mathrm{eq}}_{\alpha}(t).
    \label{DBC_mass action}
\end{align}
Combining Eq.~\eqref{DBC_mass action} and Eq.~\eqref{Force_mass action}, we obtain
\begin{align}
    F_e(t)=-\sum_{\alpha\in\mathcal{S}}\nabla_{e\alpha}\ln \frac{x_{\alpha}(t)}{x^{\mathrm{eq}}_{\alpha}(t)}.
\end{align}
We can rewrite this as $F(t)=-\nabla\psi(t)$ using $\psi_{\alpha}(t)=\ln[x_{\alpha}(t)/x^{\mathrm{eq}}_{\alpha}(t)]$. Thus, the forces are conservative at time $t$.

Second, we prove (ii)$\Rightarrow$(i). Now, there exists a potential $\psi(t)$ that satisfies $F(t)=-\nabla\psi(t)$. We take $\Xi_{\alpha}>0$ arbitrarily so that $\sum_{\alpha\in\mathcal{S}}\nabla_{\alpha e}\ln \Xi_{\alpha}=0$ holds for all $e\in\mathcal{E}$. We introduce $y$ using $\Xi_{\alpha}$ as $y_{\alpha}\coloneqq\Xi_{\alpha}\mathrm{e}^{-\psi_{\alpha}(t)+\ln x_{\alpha}(t)}$. Then, the state $y$ satisfies $J^{+}_e(y,t)=J^{-}_e(y,t)$. This is verified by
\begin{align}
    \ln\frac{J^{+}_e(y,t)}{J^{-}_e(y,t)}&=\ln\frac{\kappa_{e}^{+}(t)\prod_{\alpha\in\mathcal{S}}y_{\alpha}^{n_{\alpha e}^{+}}}{\kappa_{e}^{-}(t)\prod_{\alpha\in\mathcal{S}}y_{\alpha}^{n_{\alpha e}^{-}}}\notag\\
    &=\ln\frac{\kappa_{e}^{+}(t)}{\kappa_{e}^{-}(t)}-\sum_{\alpha\in\mathcal{S}}\nabla_{e\alpha}\ln y_{\alpha}\notag\\
    &=\ln\frac{\kappa_{e}^{+}(t)}{\kappa_{e}^{-}(t)}-\sum_{\alpha\in\mathcal{S}}\nabla_{e\alpha}\{\ln x_{\alpha}(t)-\psi_{\alpha}(t)\}\notag\\
    &=F_e(t)+(\nabla\psi(t))_{e}=0.
\end{align}
Here, we use Eq.~\eqref{Force_mass action} in the fourth transform.
We note that we can make the values of the conserved quantities~\cite{schnakenberg1976network} in $y$ the same as those in $x(t)$. In particular, we can make $y$ satisfy $\sum_{\alpha\in\mathcal{S}}y_{\alpha}=1$ by choosing $\Xi_{\alpha}=[\sum_{\alpha\in\mathcal{S}}\mathrm{e}^{-\psi_{\alpha}(t)+\ln x_{\alpha}(t)}]^{-1}$ if we consider the MJP.

\section{The general activity in the continuum limit}
\label{ap:continuum}

We consider a Brownian particle in the $n$-dimensional Euclidean space. We let $P(\bm{r})$ denote the probability density that the particle is located at $\bm{r}=(r_i)_{i=1}^{n}\in\mathbb{R}^n$. We assume that the particle is driven by a potential $V(\bm{r})$. We do not assume that the medium is isotropic or homogeneous: the mobility of the $i$-th direction at $\bm{r}$ is given by $\mu^{\mathrm{mob}}_i(\bm{r})$ and the temperature at $\bm{r}$ is given by $T(\bm{r})$. Then, the probability density evolves according to the following Fokker--Planck equation~\cite{van1988diffusion}:
\begin{align}
    \partial_tP(\bm{r})=&-\sum_{i=1}^{n}\partial_{r_i}[\mu^{\mathrm{mob}}_i(\bm{r}) [\partial_{r_i}V(\bm{r})]P(\bm{r})]\notag\\
    &+\sum_{i=1}^{n}\partial_{r_i}[\mu^{\mathrm{mob}}_i(\bm{r})\partial_{r_i}[T(\bm{r})P(\bm{r})]].
    \label{FPeq}
\end{align}

We introduce an MJP that discretizes the above system. We consider an $n$-dimensional square lattice with the lattice constant $\epsilon_L\ll 1$. We let $\bm{\alpha}=(\alpha_i)_{i=1}^{n}$ denote the coordinate of a vertex of the lattice. We also define a vector $\bm{\delta}^{i}$ as 
\begin{align}
    (\bm{\delta}^{i})_{j}\coloneqq\begin{cases}
        \epsilon_L & (j=i)\\
        0 & (j\neq i)
    \end{cases}.
\end{align}
We consider an MJP on this lattice, where the jumps occur only between adjacent vertices. Letting $R_{\bm{\alpha}\to\bm{\alpha}\pm\bm{\delta}^{i}}$ denote the rate of the jump from $\bm{\alpha}$ to $\bm{\alpha}\pm\bm{\delta}^{i}$, the time evolution of the probability distribution $x_{\bm{\alpha}}$ is given by the following linear master equation:
\begin{align}
    d_t x_{\bm{\alpha}}=&\sum_{i=1}^{n}\left(R_{\bm{\alpha}+\bm{\delta}^{i}\to\bm{\alpha}}x_{\bm{\alpha}+\bm{\delta}^{i}}-R_{\bm{\alpha}\to\bm{\alpha}+\bm{\delta}^{i}}x_{\bm{\alpha}}\right)\notag\\
    &+\sum_{i=1}^{n}\left(R_{\bm{\alpha}-\bm{\delta}^{i}\to\bm{\alpha}}x_{\bm{\alpha}-\bm{\delta}^{i}}-R_{\bm{\alpha}\to\bm{\alpha}-\bm{\delta}^{i}}x_{\bm{\alpha}}\right).
\end{align}
Here, we define the rate as
\begin{align}
    R_{\bm{\alpha}\to\bm{\alpha}\pm\bm{\delta}^{i}}\coloneqq\frac{\mathcal{A}_i(\bm{\alpha})}{\epsilon_L^2}\mp\frac{\mathcal{B}_i(\bm{\alpha})}{2\epsilon_L}+O(1),
    \label{lattice rate}
\end{align}
with
\begin{align}
    \mathcal{A}_i(\bm{\alpha})&\coloneqq\mu^{\mathrm{mob}}_i(\bm{\alpha})T(\bm{\alpha}),\\
    \mathcal{B}_i(\bm{\alpha})&\coloneqq -\mu^{\mathrm{mob}}_i(\bm{\alpha})\partial_{r_i}V(\bm{\alpha})-T(\bm{\alpha})\partial_{r_i}\mu^{\mathrm{mob}}_i(\bm{\alpha}).
\end{align}
Then, we can rewrite the linear master equation as 
\begin{align}
    d_t x_{\bm{\alpha}}=&\sum_{i=1}^{n}\left[\frac{\mathcal{B}_i(\bm{\alpha}+\bm{\delta}^i)x_{\bm{\alpha}+\bm{\delta}^i}-\mathcal{B}_{i}(\bm{\alpha}-\bm{\delta}^i)x_{\bm{\alpha}-\bm{\delta}^i}}{2\epsilon_{L}}\right]\notag\\
    &+\sum_{i=1}^{n}\left[\frac{\mathcal{A}_i(\bm{\alpha}+\bm{\delta}^i)x_{\bm{\alpha}+\bm{\delta}^i}-\mathcal{A}_i(\bm{\alpha})x_{\bm{\alpha}}}{\epsilon_L^2}\right.\notag\\
    &\left.\phantom{-\sum_{i=1}^{n}}-\frac{\mathcal{A}_i(\bm{\alpha})x_{\bm{\alpha}}-\mathcal{A}_i(\bm{\alpha}-\bm{\delta}^i)x_{\bm{\alpha}-\bm{\delta}^i}}{\epsilon_L^2}\right]+O(\epsilon_L).
    \label{fp master}
\end{align}
Here, the last term $O(\epsilon_L)$ is derived from the product of $O(1)$ term in the rate~\eqref{lattice rate} and the difference $x_{\bm{\alpha}\pm\bm{\delta}^i}-x_{\bm{\alpha}}$.
We prove that this MJP recovers the original Fokker--Planck equation~\eqref{FPeq} in the continuum limit $\epsilon_L\to 0$. Note that the probability distribution $x_{\bm{\alpha}}$ recovers the probability density $P(\bm{r})$ in the continuum limit as
\begin{align}
    P(\bm{\alpha})=\lim_{\epsilon_L\to 0}\frac{x_{\bm{\alpha}}}{\epsilon_L^n}.
    \label{prob_continuum}
\end{align}
Due to this fact, we obtain 
\begin{align}
    \partial_tP(\bm{r})=\sum_{i=1}^{n}\partial_{r_i}(\mathcal{B}_i(\bm{r})P(\bm{r}))+\sum_{i=1}^{n}\partial_{r_i}^2(\mathcal{A}_i(\bm{r})P(\bm{r})),
    \label{pre FP}
\end{align}
by deviding the both sides of equation~\eqref{fp master} by $\epsilon_L^n$, taking the limit $\epsilon_L\to 0$, and replacing $\bm{\alpha}$ with $\bm{r}$. Substituting the definition of $\mathcal{A}_i$ and $\mathcal{B}_i$, we can reduce Eq.~\eqref{pre FP} to the original Fokker--Planck equation~\eqref{FPeq}.

Using this MJP, which converges to the original continuous system, we obtain the following relation,
\begin{align}
    \lim_{\epsilon_L\to0}\epsilon_L^2\mu_m=\int_{\mathbb{R}^{n}}d\bm{r}\,P(\bm{r})\sum_{i=1}^{n}\mu^{\mathrm{mob}}_i(\bm{r})T(\bm{r}),
    \label{activity in continuum limit}
\end{align}
which indicates the behavior of the general activity in the continuum limit. Due to the Einstein relation, $\mu^{\mathrm{mob}}_i(\bm{r})T(\bm{r})$ equals the diffusion coefficient for the direction $i$ at $\bm{r}$. Thus, the generalized activity corresponds to the average of the diffusion coefficient. Note that this result is invariant for the choice of $m$. 

Let us derive the relation in Eq.~\eqref{activity in continuum limit}. In the MJP, the general activity is given by
\begin{align}
    \mu_m=\frac{1}{2}\sum_{\bm{\alpha}}&\left[\sum_{i=1}^{n}m\left(R_{\bm{\alpha}+\bm{\delta}^{i}\to\bm{\alpha}}x_{\bm{\alpha}+\bm{\delta}^{i}},R_{\bm{\alpha}\to\bm{\alpha}+\bm{\delta}^{i}}x_{\bm{\alpha}}\right)\right.\notag\\
    &\quad+\left.\sum_{i=1}^{n}m\left(R_{\bm{\alpha}-\bm{\delta}^{i}\to\bm{\alpha}}x_{\bm{\alpha}-\bm{\delta}^{i}},R_{\bm{\alpha}\to\bm{\alpha}-\bm{\delta}^{i}}x_{\bm{\alpha}}\right)\right].
    \label{lattice activity}
\end{align}
In the following, we explicitly consider the relation Eq.~\eqref{prob_continuum} as $x_{\bm{\alpha}}=\epsilon_L^nP(\bm{\alpha})+O(\epsilon_L^{n+1})$. We can obtain the asymptotic behavior of $\epsilon_L^2m(R_{\bm{\alpha}+\bm{\delta}^{i}\to\bm{\alpha}}x_{\bm{\alpha}+\bm{\delta}^{i}},R_{\bm{\alpha}\to\bm{\alpha}+\bm{\delta}^{i}}x_{\bm{\alpha}})$ as 
\begin{align}
    &\epsilon_L^2m(R_{\bm{\alpha}+\bm{\delta}^{i}\to\bm{\alpha}}x_{\bm{\alpha}+\bm{\delta}^{i}},R_{\bm{\alpha}\to\bm{\alpha}+\bm{\delta}^{i}}x_{\bm{\alpha}})\notag\\
    &=\epsilon_L^nm(\epsilon_L^2R_{\bm{\alpha}+\bm{\delta}^{i}\to\bm{\alpha}}P(\bm{\alpha}+\bm{\delta}^{i}),\epsilon_L^2R_{\bm{\alpha}\to\bm{\alpha}+\bm{\delta}^{i}}P(\bm{\alpha}))\notag\\
    &=\epsilon_L^{n}m(\mathcal{A}_i(\bm{\alpha})P(\bm{\alpha})+O(\epsilon_L),\mathcal{A}_i(\bm{\alpha})P(\bm{\alpha})+O(\epsilon_L))\notag\\
    &=\epsilon_L^{n}\mu^{\mathrm{mob}}_i(\bm{\alpha})T(\bm{\alpha})P(\bm{\alpha})+O(\epsilon_L^{n+1}).
\end{align}
Here, we use the homogeneity of $m$ in the first transform. In the same manner, we can also obtain
\begin{align}
    &\epsilon_L^2m(R_{\bm{\alpha}-\bm{\delta}^{i}\to\bm{\alpha}}x_{\bm{\alpha}-\bm{\delta}^{i}},R_{\bm{\alpha}\to\bm{\alpha}-\bm{\delta}^{i}}x_{\bm{\alpha}})\notag\\
    &=\epsilon_L^{n}\mu^{\mathrm{mob}}_i(\bm{\alpha})T(\bm{\alpha})P(\bm{\alpha})+O(\epsilon_L^{n+1}).
\end{align}
Combining Eq.~\eqref{lattice activity} and these asymptotic behaviors, we obtain
\begin{align}
    \epsilon_L^2\mu_m&=\sum_{\bm{\alpha}}\sum_{i=1}^{n}[\epsilon_L^{n}\mu^{\mathrm{mob}}_i(\bm{\alpha})T(\bm{\alpha})P(\bm{\alpha})+O(\epsilon_L^{n+1})]\notag\\
    &=\sum_{\bm{\alpha}}\epsilon_L^{n}\left[P(\bm{\alpha})\sum_{i=1}^{n}\mu^{\mathrm{mob}}_i(\bm{\alpha})T(\bm{\alpha})\right]+O(\epsilon_L).
\end{align}
This equation leads to Eq.~\eqref{activity in continuum limit} in the continuum limit $\epsilon_L\to0$.

\section{Rewriting the two conditions in Eqs.~\eqref{condition_fM'} and ~\eqref{condition_fM''}}
\label{ap:rewriting conditions}

In this appendix, we rewrite the conditions on $f_m$ in Eqs.~\eqref{condition_fM'} and ~\eqref{condition_fM''} in terms of the function defined in Eq.~\eqref{Psi_m} as
\begin{align}
    \Psi_m(u)=\frac{\mathrm{e}^u-1}{f_m(\mathrm{e}^{u})}.
\end{align}

\subsection{Rewriting the first condition~\eqref{condition_fM'}}
\label{ap:rewriting first condition}

The first condition~\eqref{condition_fM'} is equivalent to the monotonically increasingness of $\Psi_m$. It can be proved as follows: The condition~\eqref{condition_fM'} is equivalent to
\begin{align}
    \forall u,\;f_{m}'(\mathrm{e}^u)+f_{m}'(\mathrm{e}^{-u})>0.
\end{align}
This equals the positivity of $\Psi_{m}'(u)$, since the property of $f_{m}'$ in Eq.~\eqref{symmetry_fM'} leads to
\begin{align}
    \Psi_{m}'(u)&=\frac{\mathrm{e}^{u}}{f_{m}(\mathrm{e}^u)^2}\{f_m'(\mathrm{e}^u)+f_m(\mathrm{e}^u)-\mathrm{e}^uf_m'(\mathrm{e}^u)\}\notag\\
    &=\frac{\mathrm{e}^{u}}{f_{m}(\mathrm{e}^u)^2}\{f_{m}'(\mathrm{e}^u)+f_{m}'(\mathrm{e}^{-u})\}>0.
    \label{first derivative}
\end{align}
Therefore, the first condition lets us define the inverse function $\Psi_m^{-1}$. 

\subsection{Rewriting the second condition~\eqref{condition_fM''}}
\label{ap:rewriting second condition}

Under the first condition~\eqref{condition_fM'}, the other condition on $f_m$~\eqref{condition_fM''} is equivalent to the convexity of $w\Psi_m^{-1}(w)$. It can be proved as follows: Since $w\Psi_m^{-1}(w)$ is monotonically increasing on $w>0$ and even, it is enough to see the convexity of $w\Psi_m^{-1}(w)$ on $w>0$. Letting $u$ denote $\Psi_m^{-1}(w)$, we obtain
\begin{align}
    (w\Psi_m^{-1}(w))''=\frac{2\Psi_m'(u)^2-\Psi_m(u)\Psi_m''(u)}{\Psi_m'(u)^3},
    \label{inverse_derivation}
\end{align}
by direct calculation. Note that the positivity $w>0$ makes $u=\Psi_m^{-1}(w)$ positive. Since $\Psi_m'(u)$ is positive, the convexity of $w\Psi_m^{-1}(w)$ on $w>0$ is equivalent to the nonnegativity of the numerator in Eq.~\eqref{inverse_derivation} on $u>0$. We can rewrite the numerator in Eq.~\eqref{inverse_derivation} as
\begin{align}
    &2\Psi_m'(u)^2-\Psi_m(u)\Psi_m''(u)\notag\\
    &=\frac{\mathrm{e}^u(\mathrm{e}^u-1)^2f_m''(\mathrm{e}^u)+(\mathrm{e}^u+1)(f_m'(\mathrm{e}^u)+f_m'(\mathrm{e}^{-u}))}{\mathrm{e}^{-u}f_m(\mathrm{e}^u)^3}.
    \label{inverse derivation2}
\end{align}
Here, we use Eq.~\eqref{first derivative}, 
\begin{align}
    &\mathrm{e}^{-u}f_m(\mathrm{e}^{u})^3\Psi_m''(u)\notag\\&=\left[f_m(\mathrm{e}^{u})\left\{(1-3\mathrm{e}^u)f_m'(\mathrm{e}^u)-\mathrm{e}^{u}(\mathrm{e}^u-1)f''_m(\mathrm{e}^u)\right\}\right.\notag\\
    &\quad\left.+f_m(\mathrm{e}^{u})^2+2\mathrm{e}^{u}(\mathrm{e}^u-1)f_m'(\mathrm{e}^u)^2\right],
\end{align}
and the relation in Eq.~\eqref{symmetry_fM'}. We can verify the eqivalence of nonnegativity of Eq.~\eqref{inverse derivation2} on $u>0$ and the condition~\eqref{condition_fM''} by taking $r=\mathrm{e}^u$. Thus, the condition~\eqref{condition_fM''} is equivalent to the convexity of $w\Psi_m^{-1}(w)$.

\section{Availability of various means as the activity}
\label{ap:activity}
In this appendix, we show that the various means satisfy the conditions in Eqs.~\eqref{condition_fM'} and~\eqref{condition_fM''}. As a preparation, we introduce a useful rewrite of the condition~\eqref{condition_fM''},
\begin{align}
    \forall u>0,\;\left\{(\ln\Psi_m(u))'\right\}^2-(\ln\Psi_m(u))''\geq0.
    \label{reduced condition_fM''}
\end{align}
This expression follows from the fact that the condition~\eqref{condition_fM''} is equivalent to $2\Psi_m'(u)^2-\Psi_m(u)\Psi_m''(u)\geq 0$ for all $u>0$ as mentioned in Appendix~\ref{ap:rewriting second condition}. This representation further leads to Eq.~\eqref{reduced condition_fM''} because 
\begin{align}
    &2\Psi_m'(u)^2-\Psi_m(u)\Psi_m''(u)\notag\\
    &=2\left(\frac{\Psi_m'(u)}{\Psi_m(u)}\right)^2\Psi_m(u)^2-\Psi_m(u)\left(\frac{\Psi_m'(u)}{\Psi_m(u)}\Psi_m(u)\right)'\notag\\
    &=\Psi_m(u)^2\left[\left(\frac{\Psi_m'(u)}{\Psi_m(u)}\right)^2-\left(\frac{\Psi_m'(u)}{\Psi_m(u)}\right)'\right]\notag\\
    &=\Psi_m(u)^2\left[\left\{(\ln\Psi_m(u))'\right\}^2-(\ln\Psi_m(u))''\right].
\end{align}

\subsection{Stolarsky mean}
\label{ap:stolarsky}

Here, we verify that the Stolarsky mean~\eqref{Stolarsky} satisfies the conditions~\eqref{condition_fM'} and ~\eqref{condition_fM''}. The Stolarsky mean satisfies the condition~\eqref{condition_fM'} because $S_{p,q}(a,b)$ is monotonically increasing with respect to $a$ and $b$~\cite{leach1978,leach1983}. We can also prove that the Stolarsky mean satisfies the other condition~\eqref{condition_fM''} as below. 

Using $S_{p,q}$ as $m$ in Eq.~\eqref{Psi_m}, we obtain
\begin{align}
    \Psi_{S_{p,q}}(u)=2\sinh\frac{u}{2}\left[\frac{q\sinh(pu/2)}{p\sinh(qu/2)}\right]^{\frac{1}{q-p}}.
\end{align}
Here, we only need to consider the expression in the first line of Eq.~\eqref{Stolarsky} due to the continuity of the Stolarsky mean with respect to $p$ and $q$. This representation yields
\begin{align}
    &(\ln\Psi_{S_{p,q}}(u))'\notag\\
    &=\frac{1}{2}\coth\frac{u}{2}-\frac{1}{p-q}\left(\frac{p}{2}\coth\frac{pu}{2}-\frac{q}{2}\coth\frac{qu}{2}\right),
\end{align}
and
\begin{align}
    &(\ln\Psi_{S_{p,q}}(u))''\notag\\
    &=-\frac{1}{4}\csch^2\frac{u}{2}+\frac{p^2\csch^2\left(\dfrac{pu}{2}\right)-q^2\csch^2\left(\dfrac{qu}{2}\right)}{4(p-q)},
\end{align}
which rewrites the left-hand side of Eq.~\eqref{reduced condition_fM''} as
\begin{align}
    \left\{(\ln\Psi_{S_{p,q}}(u))'\right\}^2+\frac{1}{4}\csch^2\frac{u}{2}-\frac{\Theta_1\left(\dfrac{pu}{2}\right)-\Theta_1\left(\dfrac{qu}{2}\right)}{(p-q)u^2}
\end{align}
with $\Theta_1(z)\coloneqq z{}^{2}\csch^2z$, or equivalently,
\begin{align}
    &\frac{1}{4}\left(\coth^2\frac{u}{2}+\csch^2\frac{u}{2}\right)\notag\\
    &\qquad+\left[\frac{1}{p-q}\left(\frac{p}{2}\coth\frac{pu}{2}-\frac{q}{2}\coth\frac{qu}{2}\right)\right]^2\notag\\
    &\qquad\qquad-\frac{\Theta_2\left(\dfrac{pu}{2};\dfrac{u}{2}\right)-\Theta_2\left(\dfrac{qu}{2};\dfrac{u}{2}\right)}{(p-q)u^2},
\end{align}
with $\Theta_2(z;\theta)\coloneqq z{}^2\csch^2z+(2\theta\coth \theta)z\coth z$. Thus, it is enough to show that one of the following quantities is negative:
\begin{empheq}[left=\empheqlbrace]{align}
    &\frac{\Theta_1\left(\dfrac{pu}{2}\right)-\Theta_1\left(\dfrac{qu}{2}\right)}{p-q},\label{first expression}\\
    &\frac{\Theta_2\left(\dfrac{pu}{2};\dfrac{u}{2}\right)-\Theta_2\left(\dfrac{qu}{2};\dfrac{u}{2}\right)}{p-q}\label{second expression}.
\end{empheq}

In the following, we only consider $p\geq q$ since the Stolarsky mean is invariant with respect to the swapping of $p$ and $q$. We can classify $p\geq q$ into the following four cases: (i) $p\geq q\geq 0$, (ii) $p\geq0\geq q$ and $p\geq|q|$, (iii) $p\geq0\geq q$ and $|q|\geq p$, and (iv) $0\geq p\geq q$. As shown below, we can prove that Eq.~\eqref{first expression} or Eq.~\eqref{second expression} becomes negative for each case. It concludes that the Stolarsky mean satisfies the condition~\eqref{condition_fM''}. 

In the cases (i) and (ii), we can see that Eq.~\eqref{first expression} is negative because $\Theta_1(z)$ is even and monotonically decreasing on $z\geq0$. Indeed, this $z$-dependence of $\Theta_1(z)$ and the nonnegativity of $u$ let $\Theta_1(pu/2)-\Theta_1(qu/2)\leq0$ hold in these cases. The $z$-dependence of $\Theta_1(z)$ is verified by using $\coth z\geq 1/z$ on $z\geq0$ in $\partial_{z}\Theta_1(z)=-2z^2\csch^2z(\coth z-1/z)$.

In the remaining cases (iii) and (iv), we can see that Eq.~\eqref{second expression} is negative because $\Theta_2(z;\theta)$ is an even function of $z$ and monotonically increasing on $z\geq0$ for all $\theta >0$. We can easily see that this $z$-dependence of $\Theta_2(z;\theta)$ and the nonnegativity of $u$ let $\Theta_2(pu/2;u/2)-\Theta_2(qu/2;u/2)\leq0$ hold in these cases. The $z$-dependence of $\Theta_2(z;\theta)$ is verified by the following calculation for $z\geq 0, \theta>0$,
\begin{align}
    \frac{\partial_{z}\Theta_2(z;\theta)}{2z\csch^2z}&=1-z\coth z+\theta\coth\theta\left(\frac{\sinh 2z}{2z}-1\right)\notag\\
    &\geq1-z\coth z+\left(\frac{\sinh 2z}{2z}-1\right)\notag\\
    &=\frac{\sinh 2z}{2z}-z\coth z\notag\\
    &=\cosh z\left(\frac{\sinh z}{z}-\frac{z}{\sinh z}\right)\notag\\&\geq0.
\end{align}
Here, we use $\theta\coth\theta\geq1$ on $\theta>0$ and $(\sinh2z)/2z\geq1$ on $z\geq0$ to derive the first inequality.

\subsection{Contraharmonic mean}
\label{ap:contraharmonic}

Here, we verify that the contraharmonic mean $C$ satisfies the conditions~\eqref{condition_fM'} and ~\eqref{condition_fM''}. Using $C$ as $m$ in Eq.~\eqref{Psi_m}, we obtain 
\begin{align}
    \Psi_C(u)=\tanh u.
\end{align}
It immediately confirms that $C$ satisfies the condition~\eqref{condition_fM'}, since this condition is equivalent to the monotonically increasingness of $\Psi_m$ as discussed in Appendix~\ref{ap:rewriting first condition}. We also obtain
\begin{align}
    (\ln\Psi_C(u))'=\frac{1}{\sinh u\cosh u},
\end{align}
and
\begin{align}
    (\ln\Psi_C(u))''=-\frac{1}{\sinh^2 u}-\frac{1}{\cosh^2 u}.
\end{align}
These equations rewrite the left-hand side of Eq.~\eqref{reduced condition_fM''} as $2/\sinh^2u$, which is nonnegative. Hence the contraharmonic mean also satisfies the conditions~\eqref{condition_fM'} and ~\eqref{condition_fM''}.

\section{Legendre duality between current and force}
\label{ap:duality}
Let us define the dissipation functions~\cite{kobayashi2023information} as
\begin{empheq}[left=\empheqlbrace]{align}
    \Phi_m(\hat{F})&\coloneqq\sum_{e\in\mathcal{E}}\mu_{m,e}\int_0^{\hat{F}_{e}}du\,\Psi_m(u),\\
    \Phi_m^{\ast}(\hat{J})&\coloneqq\sum_{e\in\mathcal{E}}\mu_{m,e}\int_0^{\hat{J}_{e}/\mu_{m,e}}dw\,\Psi_m^{-1}(w).
    \label{dissipation function}
\end{empheq}
The dissipation functions~\eqref{dissipation function} are convex, since $\Psi_m$ and its inverse are monotonically increasing. They are also related to each other through the Legendre transform. We can verify that the Legendre transform of $\Phi_m^{\ast}(\hat{J})$ becomes $\Phi_m(\hat{F})$ as
\begin{align}
    &\sup_{I=(I_1,I_2,\cdots,I_{N_E})^{\top}}\left\{\sum_{e\in\mathcal{E}}I_e\hat{F}_e-\Phi_m^{\ast}(I)\right\}\notag\\
    &=\sum_{e\in\mathcal{E}}\mu_{m,e}\left\{\hat{F}_e\Psi_m(\hat{F}_e)-\int_{0}^{\Psi_m(\hat{F}_e)}dw\,\Psi_m^{-1}(w)\right\}\notag\\
    &=\sum_{e\in\mathcal{E}}\mu_{m,e}\int_{0}^{\hat{F}_e}du\,\Psi_m(u)=\Phi_m(\hat{F}_e),
\end{align}
where the second line is obtained from the condition that the $I_e$-derivative of the first line vanishes, and the third line is obtained by regarding $\hat{F}_e\Psi_m(\hat{F}_e)$ as the signed area of $[0,\Psi_m(\hat{F}_e)]\times[0,\hat{F}_e]$. Using the dissipation functions, we can rewrite the relations between the current and the force as
\begin{align}
    J_e=\partial_{\hat{F}_e}\Phi_m(F),\;F_e=\partial_{\hat{J}_e}\Phi_m^{\ast}(J).
\end{align}

Historically, the Legendre duality between current and force has been explored in the area of gradient flow~\cite{ambrosio2005gradient}, which is a branch of mathematics closely related to thermodynamics. First, the linear relation between current and force,
\begin{align}
    J_{e}=\mu_{L,e}F_{e},\;F_{e}=\frac{J_{e}}{\mu_{L,e}},
    \label{onsager_relation btw JF}
\end{align}
is induced by the quadratic dissipation functions,
\begin{align}
    \Phi_{L}(\hat{F})\coloneqq\dfrac{1}{2}\sum_{e\in\mathcal{E}}\mu_{L,e}\hat{F}_{e}{}^2,\;
    \Phi_L^{\ast}(\hat{J})\coloneqq\dfrac{1}{2}\sum_{e\in\mathcal{E}}\frac{\hat{J}_e{}^2}{\mu_{L,e}}\notag,
\end{align}
since $\Psi_L(u)=u$ and $\Psi^{-1}_L(w)=w$.
Here, the logarithmic mean $L$ is used as the activity.
This result was found for detailed balanced chemical systems~\cite{mielke2011gradient} and MJPs~\cite{maas2011gradient,chow2012fokker,mielke2013geodesic} as a direct generalization of the works on the Fokker--Planck equation by Otto and his collaborators~\cite{jordan1998variational,otto2001geometry}. This linear relation was extended to more general situations~\cite{kaiser2018canonical,yoshimura2023housekeeping} and has been used to reveal the connection between thermodynamics and optimal transport, decompose EPR, and derive thermodynamic trade-off relations~\cite{yoshimura2023housekeeping,van2023thermodynamic}.

In addition, the nonlinear relation with the geometric mean $G$ being the activity, 
\begin{align}
    J_{e}=2\mu_{G,e}\sinh\left(\frac{F_{e}}{2}\right),\;F_{e}=2\sinh^{-1}\left(\frac{J_{e}}{2\mu_{G,e}}\right),
    \label{frenetic_relation btw JF}
\end{align}
corresponding to the nonquadratic dissipation functions,
\begin{align}
    \Phi_{G}(\hat{F})&\coloneqq4\sum_{e\in\mathcal{E}}\mu_{G,e}\left(\cosh\frac{\hat{F}_e}{2}-1\right),\notag\\
    \Phi_{G}^{\ast}(\hat{J})&\coloneqq\sum_{e\in\mathcal{E}}\left[2\hat{J}_e\sinh^{-1}\frac{\hat{J}_e}{2\mu_{G,e}}\right.\notag\\
    &\phantom{\coloneqq\sum_{e\in\mathcal{E}}2\hat{J}_e}\left.-4\mu_{G,e}\left\{\sqrt{\frac{\hat{J}_e^2}{4\mu_{G,e}^2}+1}-1\right\}\right]\notag,
\end{align}
was discovered by considering the consistency with macroscopic fluctuation theory~\cite{mielke2014relation,mielke2017non,kaiser2018canonical}. Here, we used $\Psi_G(u)=2\sinh(u/2)$ and $\Psi^{-1}_G(w)=2\sinh^{-1}(w/2)$. This nonlinear relation is also used to bound~\cite{maes2017frenetic} and decompose the EPR~\cite{kobayashi2022hessian}.

\section{Thermodynamic trade-offs for general currents}

\subsection{The derivation of the bound for general currents}
\label{ap:generalcurrents}
The bound~\eqref{finitetimeTUR} is shown as follows. We can obtain a lower bound of the EPR by using the convexity and evenness of the function $w\Psi_m^{-1}(w)$ as
\begin{align}
    \sigma&\geq \sum_{e\in \mathcal{E}} \mu_{m,e} \left( \frac{c_eJ_e}{|c|_{\infty} \mu_{m,e}} \right)\Psi_{m}^{-1}\left(\frac{c_eJ_e}{|c|_{\infty}\mu_{m,e}}\right)\notag\\
    &=\sum_{e\in \mathcal{E}}\mu_{m}\frac{\mu_{m,e}}{\mu_{m}}\left(\frac{c_eJ_e}{|c|_{\infty}\mu_{m,e}}\right)\Psi_{m}^{-1}\left(\frac{c_eJ_e}{|c|_{\infty}\mu_{m,e}}\right)\notag\\
    &\geq\frac{\mathcal{J}_{c}}{|c|_{\infty}}\Psi_m^{-1}\left(\frac{\mathcal{J}_{c}}{|c|_{\infty}\mu_m}\right)\notag\\
    &=\frac{|\mathcal{J}_{c}|}{|c|_{\infty}}\Psi_m^{-1}\left(\frac{|\mathcal{J}_{c}|}{|c|_{\infty}\mu_m}\right).
    \label{EPR bound}
\end{align}
Here, the first line is a consequence of the inequality
\begin{align}
    \frac{J_e}{\mu_{m,e}}\Psi_{m}^{-1}\left(\frac{J_{e}}{\mu_{m,e}}\right)\geq \frac{c_eJ_e}{|c|_{\infty} \mu_{m,e}}\Psi_{m}^{-1}\left(\frac{c_eJ_e}{|c|_{\infty}\mu_{m,e}}\right),
\end{align}
where we use $-1\leq c_e/|c|_{\infty}\leq1$ and the fact that $w\Psi_m^{-1}(w)$ is an even convex function. The inequality between the second and third lines owes to Jensen's inequality with the weight $\{\mu_{m,e}/\mu_m\}_{e\in \mathcal{E}}$. In the last line, we use the evenness of $w\Psi_m^{-1}(w)$. Taking the time average of Eq.~\eqref{EPR bound}, we obtain the desired result~\eqref{finitetimeTUR}
\begin{align}
    \langle\sigma\rangle_{\tau}&\geq\frac{1}{\tau}\int_{0}^{\tau}dt\,\frac{|\mathcal{J}_{c}|}{|c|_{\infty}}\Psi_m^{-1}\left(\frac{|\mathcal{J}_{c}|}{|c|_{\infty}\mu_m}\right)\notag\\
    &={\langle\mu_m\rangle_{\tau}}\int_{0}^{\tau}dt\,\frac{\mu_m}{\tau\langle\mu_m\rangle_{\tau}}\frac{|\mathcal{J}_{c}|}{|c|_{\infty}\mu_m}\Psi_m^{-1}\left(\frac{|\mathcal{J}_{c}|}{|c|_{\infty}\mu_m}\right)\notag\\
    &\geq\frac{\langle|\mathcal{J}_{c}|\rangle_{\tau}}{|c|_{\infty}}\Psi_m^{-1}\left(\frac{\langle|\mathcal{J}_{c}|\rangle_{\tau}}{|c|_{\infty}\langle\mu_m\rangle_{\tau}}\right),
\end{align}
where we use Jensen's inequality with the weight $\{\mu_m/(\tau\langle\mu_m\rangle_{\tau})\}_{t\in[0,\tau]}$ in the last line.

\subsection{Bound for statewise observables}
\label{ap:statewisebound}

We also introduce the trade-off for the observable $\phi=(\phi_1,\cdots,\phi_{N_S})^{\top}$, which possibly depends on time.
This trade-off is a special case of the bound with a general current~\eqref{finitetimeTUR}. We use the inner product $\cip{x}{\varphi}$ between two $N_S$ dimensional vectors $x$ and $\phi$, defined as $\cip{x}{\phi}\coloneqq\sum_{\alpha\in\mathcal{S}}x_{\alpha}\phi_{\alpha}$. In the following, we call the inner product $\cip{x}{\phi}$ the \textit{expected value} of $\varphi$, since it is precisely what is called the expected value in the statistical sense in the case of the MJP.

The trade-off relation is obtained by using the speed of the time evolution measured with the observable $\phi$, which is defined as
\begin{align}
    \mathcal{V}_{\phi}\coloneqq\langle d_tx,\phi\rangle.
\end{align}
Taking $\nabla\phi$ as $c$ in Eq.~\eqref{finitetimeTUR} and using $\mathcal{J}_{\nabla\phi}=\sum_{e\in\mathcal{E}}(\nabla\phi)_{e}J_e=\langle\nabla^{\top}J,\phi\rangle=\langle d_t x,\phi\rangle$, we can easily obtain
\begin{align}
    \langle\sigma\rangle_{\tau}\geq\frac{\langle |\mathcal{V}_{\phi}|\rangle_{\tau}}{|\nabla\phi|_{\infty}}\Psi_m^{-1}\left(\frac{\langle |\mathcal{V}_{\phi}|\rangle_{\tau}}{|\nabla\phi|_{\infty}\langle\mu_m\rangle_{\tau}}\right).
    \label{finitetimebound for observable}
\end{align}
Here, we can regard $|\nabla\phi|_{\infty}$ as the inhomogeneity of the observable on the graph or CRN. This bound~\eqref{finitetimebound for observable} indicates that the larger dissipation required to realize the faster speed measured with the observable $\phi$. In contrast to the general case~\eqref{finitetimeTUR}, the bound vanishes when the system is in a steady state, since the speed $\mathcal{V}_{\phi}$ also vanishes in the steady state.

When the observable $\phi$ is time independent, the time average of the speed $\mathcal{V}_{\phi}$ is bounded by the expected value of $\phi$ at the initial time $0$ and the final time $\tau$ as
\begin{align}
    \langle |\mathcal{V}_{\phi}|\rangle_{\tau}\geq\langle \mathcal{V}_{\phi}\rangle_{\tau}\geq\frac{\cip{x(\tau)}{\phi}-\cip{x(0)}{\phi}}{\tau},
    \label{speed bound}
\end{align}
since the inhomogeneity $|\nabla\phi|_{\infty}$ is also time independent and the speed $\mathcal{V}_{\phi}$ matches the speed of the expected value of $\phi$ as $\mathcal{V}_{\phi}=d_t\cip{x}{\phi}$. The inequality~\eqref{speed bound} and the evenness of $\omega\Psi_m^{-1}(\omega)$ provide the lower bound of the right hand side in Eq.~\eqref{finitetimebound for observable},
\begin{align}
    \frac{|\cip{x(\tau)}{\phi}-\cip{x(0)}{\phi}|}{\tau|\nabla\phi|_{\infty}}\Psi_m^{-1}\left(\frac{|\cip{x(\tau)}{\phi}-\cip{x(0)}{\phi}|}{\tau|\nabla\phi|_{\infty}\langle\mu_m\rangle_{\tau}}\right),
\end{align}
which is an increasing function of $|\cip{x(\tau)}{\phi}-\cip{x(0)}{\phi}|$ and describes the trade-off between the dissipation and the magnitude of change in the expected value of $\phi$ over the finite time duration $[0,\tau]$.

\section{Kantorovich--Rubinstein duality}
\label{ap:KR dual}
We introduce another representation of the 1-Wasserstein distance, the Kantorovich--Rubinstein duality~\cite{villani2009optimal,peyre2019computational}. In the following part, we use the inner product $\cip{\cdot}{\cdot}$ introduced in Appendix~\ref{ap:statewisebound}.

This duality formula is given by the maximization problem
\begin{align}
    W_1(x^A,x^B)=\sup_{\varphi=(\varphi_{\alpha})_{\alpha\in\mathcal{S}}}\cip{x^B-x^A}{\varphi},
    \label{KR dual}
\end{align}
under the condition
\begin{align}
    \max_{e\in\mathcal{E}}|(\nabla\varphi)_e|\leq 1.
    \label{KR con}
\end{align}

The optimizer $\varphi^{\star}$ of this maximization problem [Eq.~\eqref{KR dual}] corresponds to the sign of the optimizer $U^{\star}$ of the Beckmann problem [Eq.~\eqref{Beckmann problem}]. The relation 
\begin{align}
    (\nabla\varphi^{\star})_e=\frac{U_e^{\star}}{|U_{e}^{\star}|},
    \label{KR direction}
\end{align}
holds for all $e\in\mathcal{E}^{\star}$, where $\mathcal{E}^{\star}$ is defined as $\mathcal{E}^{\star}\coloneqq\{e\mid U^{\star}_e\neq 0\}$. This is verified as below. Since the Beckmann problem and the Kantorovich--Rubinstein duality provide the same distance, we obtain 
\begin{align}
    \sum_{e\in\mathcal{E}^{\star}}|U^{\star}_e|=\sum_{e\in\mathcal{E}^{\star}}(\nabla\varphi^{\star})_eU^{\star}_e
    \label{KR direction 1}
\end{align}
by the following calculation: $\sum_{e\in\mathcal{E}^{\star}}|U^{\star}_e|=W_1(x^A,x^B)=\cip{x^B-x^A}{\varphi^{\star}}=\cip{\nabla^{\top}U^{\star}}{\varphi^{\star}}=\sum_{e\in\mathcal{E}^{\star}}(\nabla\varphi^{\star})_eU^{\star}_e$. The condition in Eq.~\eqref{KR con} also leads to
\begin{align}
    \max_{e\in\mathcal{E}}|(\nabla\varphi^{\star})_e|\leq 1.
    \label{KR direction 2}
\end{align}
The only way for $\varphi^{\star}$ to achieve both of Eq.~\eqref{KR direction 1} and Eq.~\eqref{KR direction 2} is to satisfy the desired relation~\eqref{KR direction}.

\section{Another derivation of the TSLs}
\label{ap:anotherproofTSL}

In this appendix, we derive the TSLs with the Kantorovich--Rubinstein duality in Appendix~\ref{ap:KR dual} and the trade-off relation for statewise observables in Eq.~\eqref{finitetimebound for observable}.

Based on the Kantorovich--Rubinstein duality, the speed measured with the 1-Wasserstein distance $v_1$ is also given by
\begin{align}
    v_1=\sup_{\phi=(\phi_{\alpha})_{\alpha\in\mathcal{S}} } \langle d_t x, \phi \rangle = \sup_{\phi=(\phi_{\alpha})_{\alpha\in\mathcal{S}} } \mathcal{V}_{\phi}
    \label{speed KR dual}
\end{align}
under the condition
\begin{align}
    \max_{e\in\mathcal{E}}|(\nabla\phi)_e|\leq 1.
\end{align}

Let $\phi^{\star}$ denote the optimizer of the maximization problem in Eq.~\eqref{speed KR dual}. Using $\phi^{\star}$ as the observable in Eq.~\eqref{finitetimebound for observable}, we obtain the TSL~\eqref{TSLs} as
\begin{align}
    \langle\sigma\rangle_{\tau}&\geq\frac{\langle |\mathcal{V}_{\phi^{\star}}|\rangle_{\tau}}{|\nabla\phi^{\star}|_{\infty}}\Psi_m^{-1}\left(\frac{\langle |\mathcal{V}_{\phi^{\star}}|\rangle_{\tau}}{|\nabla\phi^{\star}|_{\infty}\langle\mu_m\rangle_{\tau}}\right)\notag\\
    &=\frac{\langle v_1\rangle_{\tau}}{|\nabla\phi^{\star}|_{\infty}}\Psi_m^{-1}\left(\frac{\langle v_1\rangle_{\tau}}{|\nabla\phi^{\star}|_{\infty}\langle\mu_m\rangle_{\tau}}\right)\notag\\
    &=\langle v_1\rangle_{\tau}\Psi_m^{-1}\left(\frac{\langle v_1\rangle_{\tau}}{\langle\mu_m\rangle_{\tau}}\right).
\end{align}
Here, we use $\mathcal{V}_{\phi^{\star}}=v_1$ and $v_1\geq0$ between the first and second lines. We also use $|\nabla\phi^{\star}|_{\infty}=1$, which follows from the property of the optimizer of the Kantorovich--Rubinstein duality in Eq.~\eqref{KR direction}. 

We remark that the trade-off relation for general observable in Eq.~\eqref{finitetimebound for observable} is looser than the TSL~\eqref{TSLs}. In other words, the following inequality holds for all $\phi$:
\begin{align}
    \langle v_1\rangle_{\tau}\Psi_m^{-1}\left(\frac{\langle v_1\rangle_{\tau}}{\langle\mu_m\rangle_{\tau}}\right)\geq\frac{\langle |\mathcal{V}_{\phi}|\rangle_{\tau}}{|\nabla\phi|_{\infty}}\Psi_m^{-1}\left(\frac{\langle |\mathcal{V}_{\phi}|\rangle_{\tau}}{|\nabla\phi|_{\infty}\langle\mu_m\rangle_{\tau}}\right).
    \label{ineq TSL observable}
\end{align}
We can verify this inequality in the following way. We define a new observable $\phi'$ as $\phi'\coloneqq\phi/|\nabla\phi|_{\infty}$. This new observable satisfies $\mathcal{V}_{\phi'}=\mathcal{V_{\phi}}/|\nabla\phi|_{\infty}$. Thus, we can rewrite the right-hand side in Eq.~\eqref{ineq TSL observable} as
\begin{align}
    \langle |\mathcal{V}_{\phi'}|\rangle_{\tau}\Psi_m^{-1}\left(\frac{\langle |\mathcal{V}_{\phi'}|\rangle_{\tau}}{\langle\mu_m\rangle_{\tau}}\right).
    \label{new observable bound}
\end{align}
Since the new observable $\phi'$ satisfies $|\nabla\phi'|_{\infty}=1$, the representation of $v_1$ in Eq.~\eqref{speed KR dual} yields
\begin{align}
    v_1\geq|\mathcal{V}_{\phi'}|.
    \label{v1 is max}
\end{align}
We can obtain Eq.~\eqref{ineq TSL observable} from Eq.~\eqref{new observable bound} using the inequality~\eqref{v1 is max} and monotonically increasingness of $\omega\Psi_m^{-1}(\omega)$ on $\omega\geq0$.

\section{Derivation of the lower bounds of the transition time}
\label{ap:time bound}

Here, we derive the lower bounds of the transition time~\eqref{time bounds} from the series of TSLs in Eq.~\eqref{TSLs}. From Eq.~\eqref{TSLs}, we obtain
\begin{align}
    \frac{\Sigma_{\tau}}{l_{1,\tau}}=\frac{\langle\sigma\rangle_{\tau}}{\langle v_1\rangle_{\tau}}\geq\Psi_m^{-1}\left(\frac{\langle v_1\rangle_{\tau}}{\langle\mu_m\rangle_{\tau}}\right)=\Psi_m^{-1}\left(\frac{l_{1,\tau}}{\tau\langle\mu_m\rangle_{\tau}}\right).
\end{align}
Since $\Psi_m(u)$ increases monotonically on $u\geq0$, we obtain
\begin{align}
    \Psi_m\left(\frac{\Sigma_{\tau}}{l_{1,\tau}}\right)\geq\frac{l_{1,\tau}}{\tau\langle\mu_m\rangle_{\tau}},
\end{align}
which leads to
\begin{align}
    \frac{\Sigma_{\tau}}{l_{1,\tau}}\Psi_m\left(\frac{\Sigma_{\tau}}{l_{1,\tau}}\right)\geq\frac{\Sigma_{\tau}}{\tau\langle\mu_m\rangle_{\tau}}.
\end{align}
Considering the monotonically increasingness of $u\Psi_m(u)$ on $u\geq0$ and the triangle inequality $l_{1,\tau}\geq W_1(x(0),x(\tau))$, we obtain
\begin{align}
    \frac{\Sigma_{\tau}}{W_1(x(0),x(\tau))}&\Psi_m\left(\frac{\Sigma_{\tau}}{W_1(x(0),x(\tau))}\right)\notag\\\geq\frac{\Sigma_{\tau}}{l_{1,\tau}}&\Psi_m\left(\frac{\Sigma_{\tau}}{l_{1,\tau}}\right)\geq\frac{\Sigma_{\tau}}{\tau\langle\mu_m\rangle_{\tau}}.
\end{align}
Finally, we can derive the desired bounds~\eqref{time bounds} by dividing this equation by $\Sigma_{\tau}/\langle\mu_m\rangle_{\tau}$ and taking the reciprocal.

\section{Weaker TSLs with the total variational distance}
\label{ap:tvd}
The total variation distance,
\begin{align}
    d_{\mathrm{TV}}(x^A,x^B)\coloneqq\sum_{\alpha\in\mathcal{S}}\frac{|x^B_{\alpha}-x^A_{\alpha}|}{2},
\end{align}
provides a lower bound of the 1-Wasserstein distance as
\begin{align}
    W_1\left(x^A,x^B\right)\geq\frac{2 d_{\mathrm{TV}}\left(x^A,x^B\right)}{M_{|\nabla^{\top}|}}.
    \label{TVD bound}
\end{align}
Here, we let $M_{|\nabla^{\top}|}$ denote $\max_{e\in\mathcal{E}}\sum_{\alpha\in\mathcal{S}}|(\nabla^{\top})_{\alpha e}|$, which indicates the maximum value of the number of change in moleculars through one reaction if we consider CRNs. In the case of MJPs, the constant $M_{|\nabla^{\top}|}$ reduces to $2$, letting the inequality~\eqref{TVD bound} be the conventional form $W_1\left(x^A,x^B\right)\geq d_{\mathrm{TV}}\left(x^A,x^B\right)$. It is verified as $\sum_{\alpha\in\mathcal{S}}|(\nabla^{\top})_{\alpha e}|=|(\nabla^{\top})_{\mathrm{s}(e)e}|+|(\nabla^{\top})_{\mathrm{t}(e)e}|=|-1|+|1|=2$ because $\nabla^{\top}$ is an incidence matrix. We can obtain the inequality~\eqref{TVD bound} using the optimizer of Eq.~\eqref{Beckmann problem} $U^{\ast}$ as
\begin{align}
    2d_{\mathrm{TV}}\left(x^A,x^B\right)
    &=\sum_{\alpha\in\mathcal{S}}\left|\sum_{e\in\mathcal{E}}(\nabla^{\top})_{\alpha e}U^{\ast}_e\right|\notag\\
    &\leq\sum_{e\in\mathcal{E}}\sum_{\alpha\in\mathcal{S}}|(\nabla^{\top})_{\alpha e}|\left|U^{\ast}_e\right|\notag\\
    &\leq\left(\max_{e\in\mathcal{E}}\sum_{\alpha\in\mathcal{S}}|(\nabla^{\top})_{\alpha e}|\right)\sum_{e\in\mathcal{E}}\left|U^{\ast}_e\right|\notag\\
    &=M_{|\nabla^{\top}|} W_1\left(x^A,x^B\right),
\end{align}
where we use $x^B-x^A=\nabla^{\top}U^{\ast}$ in the first line.

We can also measure the speed of the time evolution with the total variation distance as
\begin{align}
    v_{\mathrm{TV}}\coloneqq\lim_{\varDelta t\to 0}\frac{d_{\mathrm{TV}}(x(t),x(t+\varDelta t))}{\varDelta t}=\frac{1}{2}\sum_{\alpha\in\mathcal{S}}\left|d_t x_{\alpha}(t)\right|.
\end{align}
The inequality between the 1-Wasserstein distance and the total variation distance leads to the inequality between $v_1$ and $v_{\mathrm{TV}}$,
\begin{align}
    v_1(t)\geq\frac{2 v_{\mathrm{TV}}(t)}{M_{|\nabla^{\top}|}}.
    \label{TVD speed bound}
\end{align}

The inequality between the speeds~\eqref{TVD speed bound} and the monotonicity of $\omega\Psi_m^{-1}(\omega)$ leads to the weaker TSL:
\begin{align}
    \langle\sigma\rangle_{\tau}&\geq \langle v_1\rangle_{\tau}\Psi_m^{-1}\left(\frac{\langle v_1\rangle_{\tau}}{\langle\mu_m\rangle_{\tau}}\right)\notag\\&\geq\frac{ 2\langle v_{\mathrm{TV}}\rangle_{\tau}}{M_{|\nabla^{\top}|}}\Psi_m^{-1}\left(\frac{2\langle v_{\mathrm{TV}}\rangle_{\tau}}{M_{|\nabla^{\top}|}\langle\mu_m\rangle_{\tau}}\right).
\end{align}
Similarly, we obtain
\begin{align}
    \langle\sigma\rangle_{\tau}&\geq\frac{W_1(x(0),x(\tau))}{\tau}\Psi_m^{-1}\left(\frac{W_1(x(0),x(\tau))}{\tau\langle\mu_m\rangle_{\tau}}\right)\notag\\
    &\geq\frac{2d_{\mathrm{TV}}(x(0),x(\tau))}{\tau M_{|\nabla^{\top}|}}\Psi_m^{-1}\left(\frac{2d_{\mathrm{TV}}(x(0),x(\tau))}{\tau M_{|\nabla^{\top}|}\langle\mu_m\rangle_{\tau}}\right),
\end{align}
using the inequality between the distances~\eqref{TVD bound}. Note that the triangle inequality for the total variation distance leads to
\begin{align}
    \langle\sigma\rangle_{\tau}&\geq\frac{2\langle v_{\mathrm{TV}}\rangle_{\tau}}{M_{|\nabla^{\top}|}}\Psi_m^{-1}\left(\frac{2\langle v_{\mathrm{TV}}\rangle_{\tau}}{M_{|\nabla^{\top}|}\langle\mu_m\rangle_{\tau}}\right)\notag\\
    &\geq\frac{2d_{\mathrm{TV}}(x(0),x(\tau))}{\tau M_{|\nabla^{\top}|}}\Psi_m^{-1}\left(\frac{2d_{\mathrm{TV}}(x(0),x(\tau))}{\tau M_{|\nabla^{\top}|}\langle\mu_m\rangle_{\tau}}\right).
\end{align}
Although these bounds are weaker than the one with the 1-Wasserstein distance, they allow us to treat the distribution $x$ and the topology of the graph or the CRNs separately.

\section{Derivation of the minimum dissipation formula based on the Kantorovich--Rubinstein duality}
\label{ap:minimum dissipation}

Here, we derive the minimum dissipation formula~\eqref{minimum_EP}.

First, we prove the inequality
\begin{align}
    \Sigma_{\tau}[J^+, J^-]\geq W_1(x(0),x(\tau))\Psi_m^{-1}\left(\frac{W_1(x(0),x(\tau))}{\tau M_0}\right).
    \label{minimum EP lower bound}
\end{align}
Since the TSLs are valid for general time evolution, we obtain
\begin{align}
    \Sigma_{\tau}[J^+, J^-]\geq W_1(x(0),x(\tau))\Psi_m^{-1}\left(\frac{W_1(x(0),x(\tau))}{\tau \langle\mu_m\rangle_{\tau}[J^{+}, J^-]}\right).
\end{align}
Considering Eq.~\eqref{condition_M0}, i.e., $\langle\mu_m\rangle_{\tau}[J^{+}, J^-]\leq M_0$ and the monotonically increasingness of $\Psi_m^{-1}(\omega)$ on $\omega\geq0$, we can obtain the desired inequality.

Second, we construct optimizers $J^{+\star}$ and $J^{-\star}$ that satisfy the equality of Eq.~\eqref{minimum EP lower bound}. Using the optimizers of the Beckmann problem~\eqref{Beckmann problem} and the Kantorovich--Rubinstein duality~\eqref{KR dual}, i.e., $U^{\star}$ and $\varphi^{\star}$, we define a new potential $\tilde{\varphi}$ as
\begin{align}
    \tilde{\varphi}\coloneqq-\Psi_m^{-1}\left(\frac{\sum_{e\in\mathcal{E}}|U^{\star}_e|}{\tau M_0}\right)\varphi^{\star}.
    \label{optimal potential}
\end{align}
With this potential, we define the fluxes $J^{+\star}$ and $J^{-\star}$ as
\begin{align}
    J^{+\star}_e\coloneqq
    \begin{cases}
    \dfrac{\exp\left[-(\nabla\tilde{\varphi})_{e}\right]}{\exp\left[-(\nabla\tilde{\varphi})_{e}\right]-1}\dfrac{U^{\star}_e}{\tau}& (e\in\mathcal{E}^{\star}) \\
    0& (e\in\mathcal{E}\setminus\mathcal{E}^{\star})\\
    \end{cases}
    ,
\end{align}
and
\begin{align}
    J^{-\star}_e\coloneqq
    \begin{cases}
    \dfrac{1}{\exp\left[-(\nabla\tilde{\varphi})_{e}\right]-1}\dfrac{U^{\star}_e}{\tau}& (e\in\mathcal{E}^{\star}) \\
    0& (e\in\mathcal{E}\setminus\mathcal{E}^{\star})\\
    \end{cases}
    .
\end{align}
Here, we use $\mathcal{E}^{\star}=\{e\mid U^{\star}_e\neq0\}$ introduced in Appendix~\ref{ap:KR dual}.
We can verify that the fluxes satisfy condition (ii) in Eq.~\eqref{condition_M0} as follows.  For $e\in\mathcal{E}^{\star}$, we obtain
\begin{align}
    m(J^{+\star}_e,J^{-\star}_e)
    &=\frac{U^{\star}_e}{\tau}\frac{f_m(\exp[-(\nabla\tilde{\varphi})_e])}{\exp\left[-(\nabla\tilde{\varphi})_{e}\right]-1}\notag\\
    &=\frac{U^{\star}_e}{\tau}\frac{1}{\Psi_m(-(\nabla\tilde{\varphi})_e)}\notag\\
    &=\frac{U^{\star}_e}{\tau}\left[\Psi_m\left(\Psi_m^{-1}\left(\frac{\sum_{e\in\mathcal{E}}|U^{\star}_e|}{\tau M_0}\right)(\nabla\varphi^{\star})_e\right)\right]^{-1}\notag\\
    &=\frac{U^{\star}_e}{\tau}\left[\Psi_m\left(\Psi_m^{-1}\left(\frac{\sum_{e\in\mathcal{E}}|U^{\star}_e|}{\tau M_0}\right)\frac{U_e^{\star}}{|U_e^{\star}|}\right)\right]^{-1}\notag\\
    &=\frac{U^{\star}_e}{\tau}\left[\frac{\sum_{e\in\mathcal{E}}|U^{\star}_e|}{\tau M_0}\frac{U_e^{\star}}{|U_e^{\star}|}\right]^{-1}\notag\\
    &=\frac{|U_e^{\star}|}{\sum_{e\in\mathcal{E}^{\star}}|U_e^{\star}|}M_0.
\end{align}
In the second line, we use the definition of $\Psi_m$~\eqref{Psi_m}. In the fourth line, we use the property of $\varphi^{\star}$ in Eq.~\eqref{KR direction}. In the fifth line, we also use the fact that the oddness of $\Psi_m$ leads to $\Psi_m(-\Psi_m^{-1}(\omega))=-\omega$ for all $\omega$. Finally, we use $\sum_{e\in\mathcal{E}}|U_e^{\star}|=\sum_{e\in\mathcal{E}^{\star}}|U_e^{\star}|$ in the last line. Because $m(J^{+\star}_e,J^{-\star}_e)=0$ for all $e\in\mathcal{E}\setminus\mathcal{E}^{\star}$, we obtain
\begin{align}
    \langle\mu_m\rangle_{\tau}[J^{+\star}, J^{-\star}]&=\frac{1}{\tau}\int_{0}^{\tau}dt\sum_{e\in\mathcal{E}}m(J^{+\star}_e,J^{-\star}_e)\notag\\
    &=\frac{1}{\tau}\int_{0}^{\tau}dt\sum_{e\in\mathcal{E}^{\star}}\left[\frac{|U_e^{\star}|}{\sum_{e\in\mathcal{E}^{\star}}|U_e^{\star}|}M_0\right]\notag\\
    &=M_0.
\end{align}

We can also verify that the fluxes $J^{+\star}$ and $J^{-\star}$ satisfy the equality of Eq.~\eqref{minimum EP lower bound} as
\begin{align}
    \Sigma_{\tau}[J^{+\star}, J^{-\star}]&=\int_{0}^{\tau}dt\sum_{e\in\mathcal{E}^{\star}}(J^{+\star}_e-J^{-\star}_e)\ln\frac{J^{+\star}_e}{J^{-\star}_e}\notag\\
    &=\int_{0}^{\tau}dt\sum_{e\in\mathcal{E}^{\star}}\frac{U^{\star}_e}{\tau}(-\nabla\tilde{\varphi})_e\notag\\
    &=\int_{0}^{\tau}dt\sum_{e\in\mathcal{E}^{\star}}\frac{U^{\star}_e}{\tau}\Psi_m^{-1}\left(\frac{\sum_{e\in\mathcal{E}}|U^{\star}_e|}{\tau M_0}\right)(\nabla\varphi^{\star})_e\notag\\
    &=\sum_{e\in\mathcal{E}^{\star}}U^{\star}_e\Psi_m^{-1}\left(\frac{\sum_{e\in\mathcal{E}}|U^{\star}_e|}{\tau M_0}\right)\frac{U^{\star}_e}{|U^{\star}_e|}\notag\\
    &=W_1(x(0),x(\tau))\Psi_m^{-1}\left(\frac{W_1(x(0),x(\tau))}{\tau M_0}\right).
\end{align}
Here, we use condition (iii), that is, we can regard $(J^{+\star}_e-J^{-\star}_e)\ln(J^{+\star}_e/J^{-\star}_e)=0$ for all $e\in\mathcal{E}\setminus\mathcal{E}^{\star}$, in the first line. We also use the definition of $\tilde{\varphi}$~\eqref{optimal potential} in the third line. In the fourth line, we perform the time integration and use the property of $\varphi^{\star}$ in Eq.~\eqref{KR direction}. In the last line, we use $\sum_{e\in\mathcal{E}}|U^{\star}_e|=\sum_{e\in\mathcal{E}^{\star}}|U^{\star}_e|=W_1(x(0),x(\tau))$.

We remark some properties of the optimizers $J^{+\star}$ and $J^{-\star}$. First, the current $J^{\star}\coloneqq J^{+\star}-J^{-\star}$ is independent of time. This is easily verified as
\begin{align}
    J^{\star}=\frac{U^{\star}}{\tau}.
\end{align}
Second, for all $e$ such that $J^{\star}_e\neq0$, the following equality holds:
\begin{align}
    \ln\frac{J^{+\star}_e}{J^{-\star}_e}=-(\nabla\tilde{\varphi})_e.
\end{align}
which implies that the optimizers $J^{+\star}$ and $J^{-\star}$ provide a conservative driving.

\section{Derivation of Eq.~\eqref{TSL1L ONS}}
\label{ap:derivation ONS TSL}

Let $J^{\mathrm{ONS}}$ denote the optimizer of the minimization problem in Eq.~\eqref{onsager excess}. Due to the condition in Eq.~\eqref{condition:onsager excess}, $J^{\mathrm{ONS}}$ satisfies $d_t x=\nabla^{\top}J^{\mathrm{ONS}}$. Thus, we obtain
\begin{align}
    v_1\leq\sum_{e\in\mathcal{E}}|J^{\mathrm{ONS}}_e|
    \label{speed ineq ons}
\end{align}
by the same calculation as for Eq.~\eqref{speed_ineq}.
We can also obtain the following inequality using the Cauchy--Schwarz inequality,
\begin{align}
    \mu_L\sigma^{\mathrm{ONS}}_{\mathrm{ex}}&=\left(\sum_{e\in\mathcal{E}}\mu_{L,e}\right)\left\{\sum_{e\in\mathcal{E}}\frac{(J^{\mathrm{ONS}}_e)^2}{\mu_{L,e}}\right\}\notag\\
    &=\left(\sum_{e\in\mathcal{E}}\sqrt{\mu_{L,e}}^2\right)\left\{\sum_{e\in\mathcal{E}}\left(\frac{|J^{\mathrm{ONS}}_e|}{\sqrt{\mu_{L,e}}}\right)^2\right\}\notag\\
    &\geq\left(\sum_{e\in\mathcal{E}}|J^{\mathrm{ONS}}_e|\right)^2.
    \label{ineq for onstsl}
\end{align}
These two inequalities in Eqs.~\eqref{speed ineq ons} and ~\eqref{ineq for onstsl} yeild
\begin{align}
    v_1\leq\sqrt{\mu_L\sigma^{\mathrm{ONS}}_{\mathrm{ex}}}.
\end{align}
Integrating both sides of this inequality, we obtain
\begin{align}
    \tau\langle v_1\rangle_{\tau}&\leq\int_{0}^{\tau}dt\sqrt{\mu_L\sigma^{\mathrm{ONS}}_{\mathrm{ex}}}\notag\\
    &\leq\sqrt{\int_{0}^{\tau}dt\,\mu_L}\sqrt{\int_{0}^{\tau}dt\,\sigma^{\mathrm{ONS}}_{\mathrm{ex}}}\notag\\
    &=\tau\sqrt{\langle\mu_L\rangle_{\tau}\langle\sigma^{\mathrm{ONS}}_{\mathrm{ex}}\rangle_{\tau}}.
    \label{TSLons1 pre}
\end{align}
Here, the second inequality follows from the Cauchy--Schwarz inequality. The inequality in Eq.~\eqref{TSLons1 pre} implies $\langle v_1\rangle_{\tau}^2\leq\langle\mu_L\rangle_{\tau}\langle\sigma^{\mathrm{ONS}}_{\mathrm{ex}}\rangle_{\tau}$, which equals the desired TSL in Eq.~\eqref{TSL1L ONS}.

%

\end{document}